\documentclass[a4paper,fleqn,usenatbib,useAMS]{mnras}

\usepackage{graphicx}	
\usepackage{graphics}
\usepackage{amsmath}	
\usepackage{amssymb}	
\usepackage{amsfonts}
\usepackage{multicol}        
\usepackage{bm}		
\usepackage{pdflscape}	
\usepackage{natbib}
\usepackage{epstopdf}

\usepackage[T1]{fontenc}
\usepackage{ae,aecompl}

\title[Impact of $1/f$ noise on cosmological parameters]{Impact of $1/f$ noise on cosmological parameter constraints for SKA intensity mapping}

\author[T.\,Chen, R.\,A.\,Battye,  A.\,A.\,Costa, C.\,Dickinson, S.\,E.\,Harper]{T.\,Chen,$^{1,2}$\!\thanks{E-mail:~\url{tianyue.chen@manchester.ac.uk}} R.\,A.\,Battye,$^1$ A.\,A.\,Costa,$^3$ C.\,Dickinson,$^1$  S.\,E.\,Harper$^1$
\\
$^1$Jodrell Bank Centre for Astrophysics, Alan Turing Building, School of Physics and Astronomy, The University of Manchester, \\
Oxford Road, Manchester, M13 9PL, UK.\\
$^2$SKA organisation, Jodrell Bank Observatory, Lower Withington, Macclesfield, Cheshire, SK11 9FT, UK.\\
$^3$Center for Gravitation and Cosmology, College of Physical Science and Technology, Yangzhou University, Yangzhou 225009, China}

\begin{document}
\label{firstpage}
\pagerange{\pageref{firstpage}--\pageref{lastpage}}
\maketitle


\begin{abstract}

 We investigate the impact of $1/f$ noise on cosmology for an intensity mapping survey with SKA1-MID Band\,1 and Band\,2. We  use a Fisher matrix approach to forecast constraints on cosmological parameters under the influence of $1/f$ noise, adopting a semi-empirical model from an earlier work,  which results from the residual $1/f$ noise spectrum after applying a component separation algorithm to remove smooth spectral components. Without $1/f$ noise, the projected constraints are  $4\%$ on $w_0$,  $1\%$ on $h$, $2\%$ on $b_{\rm HI}$ using Band\,1+\emph{Planck}, and $3\%$ on $w_0$,  $0.5\%$ on $h$, $2\%$ on $b_{\rm HI}$ using Band\,2+\emph{Planck}. A representative baseline $1/f$ noise degrades these constraints by a factor of $\sim1.5$ for Band\,1+\emph{Planck}, and $\sim1.2$ for Band\,2+\emph{Planck}.  On the power spectrum measurement, higher redshift and smaller scales are more affected by $1/f$ noise, with minimal contamination  comes from $z\lesssim1$ and $\ell\lesssim100$.  Subject to the specific scan strategy of the adopted $1/f$ noise model,   one prefers a correlated in frequency with  minimised  spectral slope, a low knee frequency, and a large telescope slew speed in order to reduce its impact.  
   
\end{abstract}



\section{Introduction}\label{sec:intro}

In recent times we have come to an era of precision cosmology, where CMB measurements from \emph{Planck} satellite constrain a number of the cosmological parameters with $\lesssim1\%$ accuracy \citep{planckVI18} within the standard $\Lambda$CDM model. However, for the study of dark sector which dominates in the  late time Universe, one requires observations at much lower redshifts ($z\lesssim1$) than the CMB ($z\approx1100$). This can be achieved by measuring Large-Scale-Structures (LSS) of the Universe. The Baryon Acoustic Oscillation (BAO) feature, for example, is a key cosmological probe, which is the imprint left by acoustic waves in the early Universe and has a known scale of $\approx150$\,Mpc    \citep[e.g.,][]{aab+14}. By measuring the  BAO scale at several redshifts, one can use it as a \emph{standard ruler} to deduce the time evolution of the Universe, and in particular, the evolution of the dark sector \citep[e.g.,][]{bull16}. 

Conventionally, the measurement of BAO is achieved through optical galaxy surveys that detect individual galaxies with high resolution optical fibres. However, an alternative approach at radio wavelength is through the intensity mapping (IM) technique.  The concept is that it maps a single emission line from multiple unresolved galaxies, each of which is below the detection limit, but resolves large-scale-structures by measuring intensity fluctuations over cosmological distances \citep{bdw04, pbp06}.  In addition to pixel-to-pixel fluctuations,  the  observing frequencies of radio IM surveys provide accurate redshift information, mapping the Universe in three dimensions \citep[e.g.,][]{wme+13, kvl+17, bbg+19}.  

Neutral hydrogen (HI) remains in dense gas clouds hosted predominantly within galaxies, and is thus a good tracer of galaxy densities  that reveals the matter density fluctuations in the Universe \citep[e.g.,][]{mmr97}. The 21\,cm emission line (also known as HI line) comes from the \emph{spin-flip} transition of electrons in neutral hydrogen, and is a good tracer of mass with minimal bias \cite[e.g.,][]{pcr15}.  

Many HI IM experiments have been proposed, with the first detection made by the GBT team at $z\approx0.8$ through the cross-correlation between HI IM maps and optical data \citep{cpb+10, msb+13}.  Some HI IM experiments propose to use a single dish operating at lower redshifts ($z<$1), such as GBT \citep{cpb+10}, BINGO \citep{bbd+13} and FAST \citep{nlj+11, bmb+16}.   Others propose to use an interferometer,  such as  TIANLAI \citep{chen12}, CHIME \citep{baa+14}, HIRAX \citep{nbb+16}, HERA \citep{dpa+17}, MWA \citep{bck+13}, LOFAR \citep{hwg+13}, PAPER \citep{pbf+10} and LWA \citep{eam+18}.  The upcoming Square Kilometre Array (SKA) is promising in terms of conducting HI IM surveys, operating in interferometry mode for SKA-LOW, and single dish (total power) mode for SKA-MID  \citep{sba+15, bfp+15, skaredbook18}. %

The success of an intensity mapping experiment will rely primarily on two aspects: i) the effective removal of Galactic foreground contamination; ii) the  control of the instrumental noise and systematic errors.  The Galactic foreground emission can be $\sim10^4$ times stronger than the HI signal so that an effective component separation method must be applied to properly reconstruct HI signal from foreground contamination \citep[e.g.,][]{wab+14, abf+15, ord16}. Systematics is another challenge which contaminates HI signal in two ways: i) it mimics or obscures HI signals; ii) it complicates the structure of Galactic foregrounds, making it difficult for component separation to properly subtract HI signals. For example, both \cite{smb+13} and \cite{pyk+17}, using GBT and LOFAR for HI IM respectively,  found that mis-calibration is one of the main limiting factors for detecting HI signals. Radio-frequency interference (RFI), such as mobile phones and satellite,  is another challenge for HI IM \citep{cpb+10,smb+13, hd18}.  

For a single dish radio telescope, significant contamination will come from the receiver $1/f$ noise, caused by gain fluctuations, $\Delta G(t)/G$,  in the receiver system \citep{Nyquist1928} resulting from  ambient temperature changes, transistor quantum fluctuations, and power voltage variations. The term ``$1/f$ noise'' originates from the shape of the power spectrum that increases approximately inversely with frequency. It can  dominate Gaussian thermal noise and contaminate HI signal.  In the observed map, $1/f$ noise will introduce stripes along the scan direction \citep{bdb+15}. Therefore, for detecting weak HI signals with IM, care must be taken to mitigate $1/f$ noise \citep{hdb+18}. 

In this work, we focus on quantifying the impact of $1/f$ noise on cosmological parameter constraints using SKA IM. We adopt the semi-empirical $1/f$ noise model from \cite{hdb+18}, and add it into a Fisher matrix analysis to project constraints on cosmological parameters, subject to the specific scan strategy and component separation method used in \cite{hdb+18}. Sect.\,\ref{datasec} introduces the SKA IM survey parameters and Sect.\,\ref{anasec} gives the formulae for power spectra and Fisher matrix calculation. Sect.\,\ref{sec:thermalonly} and Sect.\,\ref{sec:with1f} present the projected constraints with and without $1/f$ noise respectively, and in  Sect.\,\ref{dissec} we draw conclusions.


\section{The survey}\label{datasec}

The IM facility assumed in our analysis is SKA. SKA will be delivered in two phases, with SKA1 currently under construction, and the configuration of SKA2 to be decided.  SKA1 will comprise two telescopes - SKA1-MID and SKA1-LOW. SKA1-MID is a dish array based in South Africa, observing between 0.35\,GHz$-$1.75\,GHz. SKA1-LOW is located in western Australia, observing between 0.05\,GHz$-$0.35\,GHz \citep{skaredbook18}. For the purpose of  our study, we will focus on using SKA1-MID since it observes LSS at low redshifts ($z<3$) where the dark sector dominates.

Following \cite{skaredbook18}, SKA1-MID consists of 64$\times$13.5\,m MeerKAT dishes and 133$\times$15\,m SKA1 dishes. For simplicity, we assume all 197 dishes have the same dish diameter of 15\,m in our analysis, which will not lead to significant errors.  The SKA1-MID will operate in two frequency bands, with Band\,1 observing at 350\,MHz$-$1050\,MHz and Band\,2 observing at $950$\,MHz$-$1750\,MHz.  \cite{sba+15} and \cite{bfp+15} argued that compared to interferometric  mode, SKA1-MID operating in the single dish (auto-correlation) mode benefits from a better sensitivity to HI at BAO scales,  and an increased HI surface brightness temperature sensitivity. Therefore, we  primarily concentrate on  SKA1-MID Band\,1 since it is the main band for SKA IM \citep{skaredbook18}, but we will also consider SKA1-MID Band\,2 at $950$\,MHz$-$1410\,MHz, with both operating in single dish (total power) mode. A frequency channel width of 20\,MHz is assumed for both bands, which is the same as in \cite{hdb+18} for consistency. This gives a total number of 35 and 23 frequency channels for Band\,1 and Band\,2 respectively. We refer to Sect.\,\ref{sec:nbin} for more discussions on the choice of total  number of frequency channels. 

The receiver will have dual polarisation, and the beam full-width half-maximum (FWHM) is calculated at the medium  frequency of each band as
\begin{equation}
\theta_{\rm FWHM} = 1.2\frac{\lambda}{D_{\rm dish}}\,,
\end{equation}
which gives $\theta_{\rm FWHM} = 1.77^{\circ}$  for Band\,1 at $\nu_{\rm med} = 700$\,MHz, and $\theta_{\rm FWHM} = 1.05^{\circ}$ for Band\,2 at  $\nu_{\rm med} = 1180$\,MHz respectively. 

We follow \cite{skaredbook18}  to calculate the system temperature of SKA1-MID  as 
\begin{equation}
  T_{\rm sys} = T_{\rm rx} + T_{\rm spl} + T_{\rm CMB} + T_{\rm gal},
  \label{tsys}
\end{equation}
where $T_{\rm spl}\approx3$\,K is the contribution from ``spill-over'', and $T_{\rm CMB}\approx2.73$\,K is the CMB temperature. The contribution from our own Galaxy scales with frequency as
\begin{equation}
 T_{\rm gal} = 25\mathrm{\,K}(408\,\mathrm{MHz}/\nu)^{2.75}\,.
  \label{tgal}
\end{equation}
The receiver noise temperature $T_{\rm rx}$ is modelled using
\begin{equation}
  T_{\rm rx} = 15\,\mathrm{K}+ 30\,\mathrm{K}\left(\frac{\nu}{\mathrm{GHZ}}-0.75\right)^2\,
  \label{trxb1}
 \end{equation}
for Band\,1, and $T_{\rm rx} = 7.5$\,K for Band\,2. We further simplify the calculation for Band\,2 by assuming a constant Galactic contribution  of $T_{\rm gal} \approx 1.3$\,K as it is subdominant at high frequencies, and yield a constant overall system temperature of $T_{\rm sys} = 15$\,K  for Band\,2.

We follow \cite{skaredbook18} by assuming a 10000 hours integration time for the IM surveys with both bands, and a 20000\,deg$^2$ and 5000\,deg$^2$ sky coverage for Band\,1 and Band\,2 respectively. The instrumental and observing parameters of SKA1-MID Band\,1 and Band\,2 are summarised in Table\,\ref{skainput}.

\begin{table} 
\centering
\begin{tabular}{l|l|l}
\hline
\hline
Instrumental parameters & Band\,1 & Band\,2 \\
\hline
Dish diameter,  $D_{\rm dish}$ (m) & \multicolumn{2}{c}{15} \\
No. beams, $n_{\rm beam}$ (dual pol.)  & \multicolumn{2}{c}{1$\times$2} \\
No. dishes, $n_{\rm t}$  & \multicolumn{2}{c}{197} \\
Integration time,  $t_{\rm obs}$ (hrs) &  \multicolumn{2}{c}{10000}\\
Channel width, $\delta\nu$ (MHz)  & \multicolumn{2}{c}{20}     \\
Beam resolution, $\theta_{\rm FWHM}$ (deg) & 1.77 & 1.05 \\ 
System temperature, $T_{\rm sys}$ (K) & Equ.\,\ref{tsys}\&\ref{trxb1} & 15 \\
Frequency range, $\Delta\nu$ (MHz) & [350, 1050] & [950, 1410] \\
No. channels, $N_{\rm bin}$ & 35 & 23 \\
Redshift range, $[z_{\rm min}, z_{\rm max}]$ & [0.35, 3] & [0.01, 0.5] \\
Survey coverage, $\Omega_{\rm sur}$ (deg$^2$)  & 20000 & 5000 \\
\hline
$1/f$ noise parameters & & \\
\hline
Slew speed, $v_{\rm t}$ (deg\,s$^{-1}$) & \multicolumn{2}{c}{[0.5, \textbf{1}, 2]}\\
Knee frequency, $f_{\rm knee}$ (Hz) & \multicolumn{2}{c}{[0.01, 0.1, 0.5, \textbf{1}, 5, 10]} \\
Spectral index, $\alpha$ & \multicolumn{2}{c}{[\textbf{1}, 2]} \\
Correlation index, $\beta$ & \multicolumn{2}{c}{[0.25, \textbf{0.5}, 0.75, 1]} \\
\hline
\hline
\end{tabular}
\caption{The instrumental and observing parameters of SKA1-MID Band\,1 and Band\,2 \protect\citep{skaredbook18}. The properties of  $1/f$ noise model used in our analysis are listed in the bottom of the table, with the default baseline values in bold. }
\label{skainput}
\end{table}


\section{Fisher forecast formalism}\label{anasec}
We project constraints on cosmological parameters using the Fisher matrix technique \citep{fisher20}, which is a quick and effective way to obtain cosmological forecasts. It  assumes all parameters are Gaussian-distributed and no additional uncertainties unless they are incorporated. Forecasts from Fisher method give the optimal results expected from the upcoming experiment, and can guide  the instrumental requirements. In this section, we present the HI power spectrum formula, noise models,  Fisher matrix formula, and priors used in our calculations.
\subsection{HI power spectrum}
We adopt the dimensionless 2D angular power spectrum of HI signal without multiplication by the average HI brightness temperature, calculated by \citep[e.g.,][]{bd11, bbd+13} 
\begin{equation}
\begin{split}
C^{\rm HI}_{\ell}(z_i,z_j)&=\left(\frac{2}{\pi}\right)\int dz \left(W_i(z)D(z)\right) dz'\left(W_j(z')D(z')\right) \\
 &\times \int dk k^{2}P_{\rm m}(k,z=0) \left[b_{\rm HI}j_{\ell}(k\chi)-f(z)j^{\prime \prime}_{\ell}(k\chi) \right]\\
 & \times \left[b_{\rm HI}j_{\ell}(k\chi')-f(z')j^{\prime \prime}_{\ell}(k\chi') \right]\,,
\label{eq:Cell2}
\end{split}
\end{equation}
where the window function $W_i(z)$ of each bin is  centred at redshift $z_i$ with a bin width of $\Delta z$ such that
\begin{equation}
W_i(z) = \begin{cases}
  \frac{1}{\Delta z}, & z_i-\frac{\Delta z}{2}\leq z \leq z_i+\frac{\Delta z}{2}\,,\\
  0, & \text{otherwise}\,.
\end{cases}
\end{equation}
The growth factor $D(z)$ is calculated from the system of equations \citep{dod03}
\begin{equation}
 \begin{split}
D''(a) & =  -\left(\frac{3}{2}+\frac{1}{2}\frac{d\left(\mathrm{log}E(a)\right)}{da}\right)D'(a)+ \frac{3\Omega_{\rm m}}{2a^5E(a)}D(a)\,,\\
E(a) & =  \frac{H(a)}{H_0}\,,
\end{split}
\end{equation}
and the growth rate $f(z)$ is the derivative of $D(z)$ such that
\begin{equation}
f(a) = \frac{d\,\left(\mathrm{log}D(a)\right)}{d\,\mathrm{log}a}\,,
\end{equation}
which encodes the \emph{redshift-space-distortion} effect (RSD) caused by the peculiar motion of galaxies \citep[e.g.,][]{hamilton98}. The underlying matter power spectrum $P_{\rm m}$ at each wavenumber $k$ and the current redshift $z = 0$ is computed using the \textsc{camb} software \citep{aa11,hlh+12}. We assume a redshift- and scale-independent HI bias at the fiducial value of $b_{\rm HI} = 1$. The multipole  $\ell$ is related to the wavenumber $k$ through the Bessel function $j_{\ell}(k\chi)$ and its second derivative $j^{\prime\prime}_{\ell}(k\chi)$, where $\chi$ is the comoving distance defined by
\begin{equation}
\chi(z)=\int^{z}_{0}\frac{cdz'}{H(z')}\,. 
\label{eq:chi}
\end{equation}

Fig.\,\ref{fig:cov1f} shows the HI power spectrum (\emph{red}) at $z = 0.5$ (\emph{solid}) and  $z = 2$ (\emph{dashed}) respectively, illustrating that the amplitude of the power spectrum decreases towards higher redshift.  The evolution of the HI power spectrum with redshift depends on the growth of matter structure, which is sensitive to the Dark Energy equation of state parameter $w$ at low redshift \citep[e.g.,][]{odb+18}. This is why HI IM at low redshift is a good probe of the dark sector.

\subsection{Thermal noise power spectrum}
Thermal noise defines the fundamental sensitivity of the instrument. It is the voltages generated by thermal agitations in the resistive components of the receiver.  Thermal noise is calculated by the radiometer equation \citep{wrh09}
\begin{equation}
  \sigma_{\rm T} = \frac{T_{\rm sys}}{\sqrt{t_{\rm pix}\delta\nu}}\,,
\label{equ:radiometer}
\end{equation}
where $T_{\rm sys}$ is the total system temperature introduced in Equ.\,\ref{tsys}, $\delta\nu$ is the frequency channel width, and $t_{\rm pix}$ is the integration time per pixel such that 
\begin{equation}
t_{\rm pix}=t_{\rm obs}\frac{n_{\rm beam}n_{\rm t}\Omega_{\rm pix}}{\Omega_{\rm sur}}.
\end{equation}
where $\Omega_{\rm pix} \propto \theta^2_{\rm FWHM}$ is the pixel area, and other parameters are taken from Table\,\ref{skainput}.  The angular power spectrum of thermal noise is 
\begin{equation}
N_{\ell}(z_i,z_j)= \left(\frac{4 \pi}{N_{\rm pix}} \right)\sigma_{\rm T,i}\sigma_{\rm T,j}\,,
\label{noise}
\end{equation}
where $N_{\rm pix}$ is the number of pixels in the map,  and $\sigma_{\rm T,i}$ is the thermal noise level for each frequency channel $i$ given by Equ.\,\ref{equ:radiometer}. The dimensionless thermal noise power spectrum is calculated by dividing $\sigma_{\rm T,i}$ by the  mean brightness temperature of 21\,cm signal at each frequency channel, $\bar{T} (z)$, with \citep{bbd+13}
\begin{equation}
\bar{T}(z)=180 \, \Omega_{\rm HI}h  \frac{(1+z)^2}{E(z)} \, {\rm mK}\,,
\end{equation}
where $\Omega_{\rm HI}$ is the density of 21\,cm signal relative to the present-day critical density, and we assume a constant $\Omega_{\rm HI} = 6.2\times10^{-4}$ as measured by \cite{smb+13} using GBT at $z\sim0.8$. 

One also needs to apply the beam correction $b_{\ell}(z_i)$ at  each frequency channel $\nu_i$ such  that
\begin{equation} 
b_{\ell}(z_i) = \exp\left[-\frac{1}{2}\ell^{2} \sigma^2_{b, i}\right]\,,
\label{bll0}
\end{equation}
where $\sigma_{b, i}=\theta_{\rm B}(z_i)/\sqrt{8\ln 2}$ \citep[e.g.,][]{bfp+15}, and
\begin{equation}
\theta_{\rm B}(z_i)=\theta_{\rm FWHM}(\nu_{\rm mid})\frac{\nu_{\rm mid}}{\nu_i}\,,
\end{equation}
with $\nu_{\rm mid}$ being the middle frequency of the survey. The effect of the beam on the power spectrum is to reduce the signal by a factor of $b_{\ell}^2$, which can be thought of as a increase in the noise by a factor of 
\begin{equation} 
B_{\ell}(z_i, z_j) = \exp\left[\ell^{2} \sigma_{\rm b,i}\sigma_{\rm b,j}\right]\,.
\end{equation}
We will assume the thermal noise is uncorrelated in frequency such that $N_{\ell}(z_i, z_j) = 0$ for $z_i \neq z_j$.

\begin{figure}
\centering
\includegraphics[width=0.95\hsize]{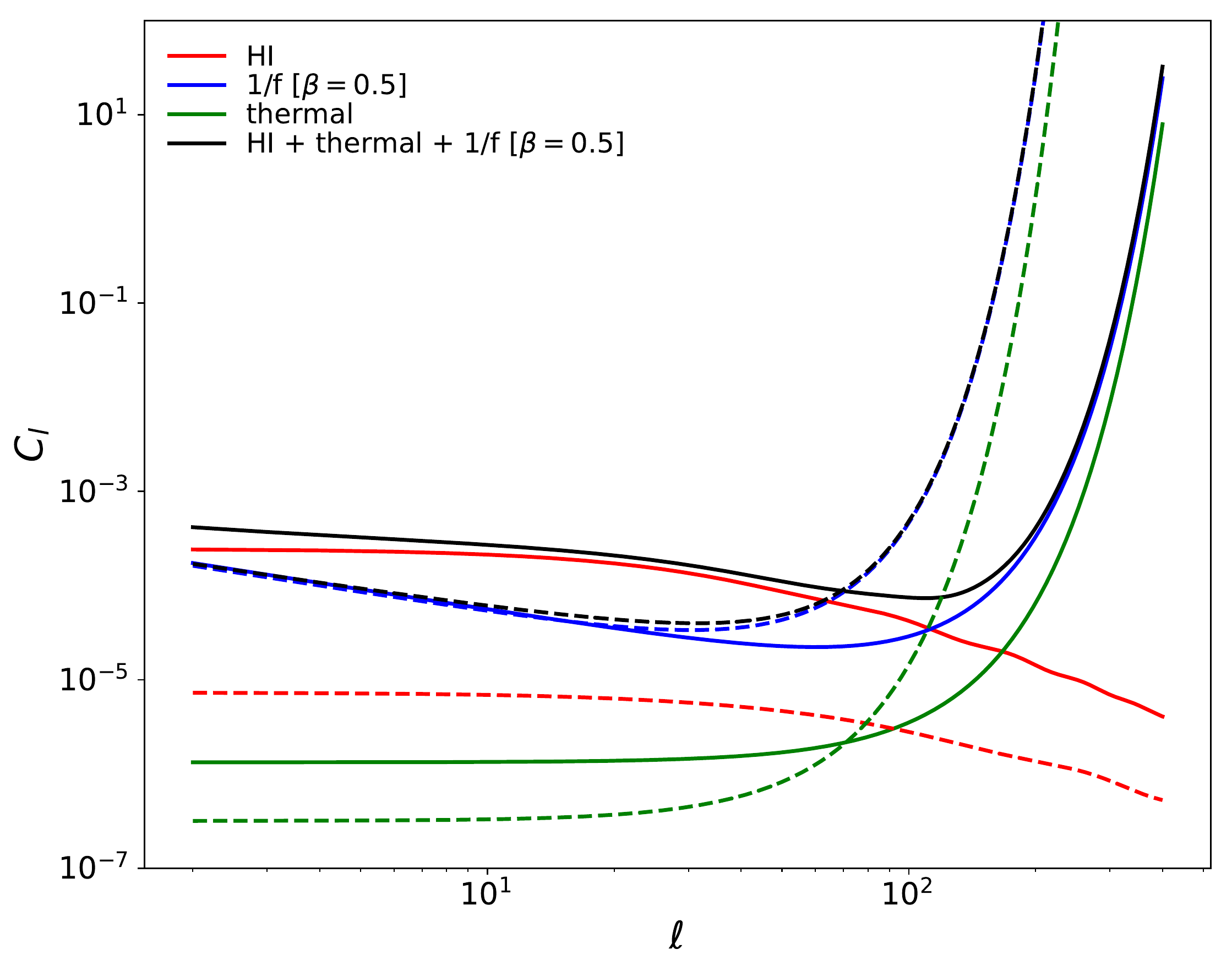}
\caption{The HI power spectrum (\emph{red}), thermal noise (\emph{green}),   $1/f$ noise (\emph{blue}) and total power spectrum (\emph{black}) for SKA1-MID Band\,1 at the redshift of $z = 0.5$ (\emph{solid}) and $z = 2$ (\emph{dashed}).  The $1/f$ noise is calculated for the baseline set up of [$\beta = 0.5$, $\alpha = 1,\, f_{\rm knee} = 1\,\mathrm{Hz},\,v_{\rm t} = 1\,\mathrm{deg\,s^{-1}}$]. Both thermal noise and $1/f$ noise have the beam correction applied. }
\label{fig:cov1f}
\end{figure}

Fig.\,\ref{fig:cov1f} shows the expected thermal noise (\emph{green}) at $z = 0.5$ (\emph{solid}) and  $z = 2$ (\emph{dashed}) respectively for SKA1-MID Band\,1. In both cases, the thermal noise will be well below the HI signal (\emph{red}) at most scales below $\ell\sim100$, enabling the detection of the signal. Above $\ell\sim100$, the thermal noise increases exponentially and surpasses the signal due to the effect of the beam at small angular scales (hight $\ell$). 

\subsection{$1/f$ noise power spectrum} \label{sec:1fintro}
As introduced in Sect.\,\ref{sec:intro}, $1/f$ noise is induced by  gain fluctuations in the receiver system and  is independent of thermal noise. For a receiver system contaminated by both, overall power spectral density (PSD) is the quadratic addition of the two components such that \citep[e.g.,][]{smb+02, bdb+15, hdb+18}
\begin{equation}
\mathrm{PSD}(f) = \sigma^2_{\rm T}\left[1+\left(\frac{f_{\rm knee}}{f}\right)^{\alpha}\right]\,,
\label{psdtot}
\end{equation}
where $\sigma_{\rm T}$ is the thermal noise level, with the first term in the bracket being the contribution from thermal noise and the second power-law term arising from $1/f$ noise.  The knee frequency, $f_{\rm knee}$, is the frequency where $1/f$ noise has the same amplitude as thermal noise. The spectral index of $1/f$ noise, $\alpha\approx1-2$,  is always positive so that $1/f$ component has more  power on longer timescales.

\cite{hdb+18} modified Equ.\,\ref{psdtot} to  take into account correlations in the  frequency  direction such that
\begin{equation}
  \mathrm{PSD}(f, \omega) = \sigma^2_{\rm T}\left[1+\left(\frac{f_{\rm knee}}{f}\right)^{\alpha}\left(\frac{\omega_0}{\omega}\right)^{\frac{1-\beta}{\beta}}\right]\,,
  \label{psd1f}
\end{equation}
where $\omega$ is $1/\nu$. 

For $N$ frequency channels and a channel width of $\delta\nu$, the values of $\omega$ range from the smallest, $\omega_0 = \frac{1}{N\delta\nu} = \frac{1}{\Delta\nu}$, to the largest, $\omega_{N-1} = \frac{1}{\delta\nu}$.   The correlation index, $\beta$,  describes the correlation of $1/f$ noise in frequency, and has a value between 0 and 1. For $\beta = 0$, the $1/f$ noise is completely correlated across all frequency channels and for $\beta = 1$, the $1/f$ noise is completely uncorrelated. 

The correlation of $1/f$ noise in frequency space, $G(\nu)$, is given by the discrete inverse  Fourier transform of the correlation in wavenumber space, such that
\begin{equation}
G(\nu_k) = \frac{1}{N}\sum^{N-1}_{n=0}\left(\frac{\omega_0}{\omega_n}\right)^{\frac{1-\beta}{\beta}}e^{\frac{2\pi i}{N}kn}\,,\;k\in[0, N-1]
\label{equ:gf}
\end{equation}
which quantifies the correlation of the first frequency channel with other channels. In order to get the covariance matrix $G(\nu_i, \nu_j)$ of $1/f$ noise, we construct a Toeplitz matrix from $G(\nu)$, which has constant descending diagonals from left to right \citep[e.g.,][]{gv96}.  The Toeplitz matrix has been used in, e.g., the \emph{Planck} CMB map-making analysis to describe the $1/f$ noise covariance matrix \citep{abb+07}. In our case, the Toeplitz matrix  $G(\nu_i, \nu_j)$ is   
\begin{equation} G(\nu_i, \nu_j) = \setlength\arraycolsep{1pt}\begin{bmatrix}
G(\nu_1) & G(\nu_2)  & \dots & G(\nu_{n-1}) & G(\nu_n) \\[10pt]
G(\nu_2) & G(\nu_1)  & \dots & G(\nu_{n-2}) & G(\nu_{n-1}) \\[10pt]
\vdots & \vdots  & \ddots & \vdots & \vdots \\[10pt]
G(\nu_{n-1}) & G(\nu_{n-2}) & \dots & G(\nu_1) & G(\nu_2)\\[10pt]
G(\nu_n) & G(\nu_{n-1})  & \dots & G(\nu_2) & G(\nu_1)
\end{bmatrix}\,.\label{equ:cov1f}\end{equation}
If the $1/f$ noise is completely correlated in frequency space, Equ.\,\ref{equ:cov1f} is a matrix of ones and for completely uncorrelated $1/f$ noise, Equ.\,\ref{equ:cov1f} is an identity matrix.

\cite{hdb+18} provided a semi-empirical angular power spectrum of the residual $1/f$ noise level in the reconstructed HI power spectrum after applying component separation technique. A high-pass filter was applied to the time-ordered-data (TOD) in their simulation to remove correlations on very long-time scales, and we have tested that the bias on the mean HI signal due to this filter has negligible impact on our results since large scales are dominated by the cosmic variance.  Note that if it is necessary to filter TOD in order to make the mean unbiased, extra corrections may become required.  In particular, the model is specific to the scan strategy and component separation method used in their analysis, and may not be valid if things are done differently. Here we adopt their model where for completely uncorrelated $1/f$ noise,  the residual angular power spectrum, $F_{\ell}(\nu, \beta = 1)$, is given by
\begin{equation}\label{eqn:emp1f} 
\begin{split}
\log_{10}\left[\frac{F_{\ell}(\nu, \beta = 1)}{\mu\rm K^2}\right] = \log_{10}\left(\frac{A}{\mu\rm K^2}\right) + a\left[\alpha - 1\right] & \\
 + b \sqrt{\alpha} \log_{10}\left( \frac{v_{\rm t}}{\mathrm{deg\,s}^{-1}} \right)-\sqrt{c\alpha} \log_{10}(\ell) & \,,
\end{split}
\end{equation}
where $\alpha$ is the $1/f$ spectral index, $v_{\rm t}$ is the telescope slew speed, and the best fitted values of $a$, $b$ and $c$ are 1.5, -1.5 and 0.5 respectively. The amplitude $A$ is parameterised by
\begin{equation}\label{eqn:emp1}
\begin{split}
A = \left(\frac{T_{\rm sys}}{21 \rm K}\right)\left(\frac{f_{\rm knee}}{1\,\text{Hz}}\right)^{\alpha}\left(\frac{\delta \nu}{20\,\text{MHz}}\right)^{-1}\left(\frac{n_{\rm t}}{200}\right)^{-1} & \\
\left(\frac{T_\text{obs}}{30\,\text{days}}\right)^{-1}\left(\frac{\Omega_{\rm sur}}{20500\,\mathrm{deg}^2}\right)10^2\mu\text{K}^2 & \,,
\end{split}
\end{equation}
where $T_{\rm sys}$ is the system temperature, $f_{\rm knee}$ is the knee frequency, $\delta \nu$ is the frequency channel width, $n_t$ is the number of telescopes, $T_\text{obs}$ is the integration time in unit of days, and $\Omega_{\rm sur}$ is the survey sky coverage. The frequency-dependence of $F_{\ell}(\nu, \beta = 1)$ comes from the  frequency-dependent $T_{\rm sys}$ calculated using Equ.\,\ref{tsys}. 

For a given value of the correlation index $\beta$, $F_{\ell}(\nu, \beta)$ is related to $F_{\ell}(\nu, \beta = 1)$ through
\begin{equation}
\frac{F_{\ell}(\nu, \beta)}{F_{\ell}(\nu, \beta = 1)} = d\sin(2\pi\beta) + \beta\,,
\label{equ:1fbeta}
\end{equation}
with $d = -0.16$. By taking the frequency correlations  into account, the $1/f$ noise for each frequency channel is the product of the $1/f$ noise covariance matrix $G(\nu_i, \nu_j)$ and the angular power spectrum $F_{\ell}(\nu, \beta)$ such that
\begin{equation}
F_{\ell}(\nu_i, \nu_j, \beta) = \mathrm{G}(\nu_i, \nu_j)\sqrt{F_{\ell}(\nu_i, \beta)F_{\ell}(\nu_j, \beta)}\,.
\end{equation}

Throughout the paper, unless otherwise stated, we adopt the same  baseline setup of $1/f$ noise at [$\beta = 0.5$, $\alpha = 1$, $f_{\rm knee} = 1\,\mathrm{Hz}$, $v_{\rm t} = 1\,\mathrm{deg\,s^{-1}}$] as in \cite{hdb+18}. We vary the value of each $1/f$ noise parameter in Sect.\,\ref{sec:basepar}, to investigate their impact on cosmological parameters. The ranges of these parameters are listed in Table\,\ref{skainput}.  Fig.\,\ref{fig:cov1f} shows the $1/f$ noise (\emph{blue}) using the baseline setup. For $z = 0.5$ (\emph{solid}), the $1/f$ noise is below the HI signal (\emph{red}) until $\ell\sim100$, enabling a detection of the signal,  but nevertheless higher than the thermal noise (\emph{green}) over all scales. The total observed power spectrum (\emph{solid black}) in this case is dominated primarily by HI signal below $\ell\sim100$, and by $1/f$ noise above $\ell\sim100$ due to the effect of the beam.   At $z = 2$ (\emph{dashed}), the $1/f$ noise is significantly higher than both HI signal (\emph{red}) and thermal noise (\emph{green}) at all scales, where no detection of signal can be made. The total observed power spectrum (\emph{dashed black}) in this case is always dominated by $1/f$ noise. Based on Fig.\,\ref{fig:cov1f}, $1/f$ noise could have a big impact on the HI signal detection, and thus cannot be ignored.

\subsection{Fisher matrix and cosmological parameters} \label{sec:fm}
The Fisher matrix for projecting cosmological parameter constraints  is constructed  by \citep[e.g.,][]{dod03, acg+12, hc12}
\begin{equation}
M_{ij} = \sum_{\ell=2}^{\ell_{\rm max}}\sum_{XX',YY'}\frac{\partial C_{\ell}^{XX'}}{\partial \theta_i}\left[\text{Cov}(XX', YY')\right]^{-1}_{\ell}\frac{\partial C_{\ell}^{YY'}}{\partial \theta_j},
\label{equfish}
\end{equation}
where $C_{\ell}^{XX'}$ and $C_{\ell}^{YY'}$ are the HI power spectra (Equ.\,\ref{eq:Cell2}) at different redshift bins denoted by $X$, $X'$, $Y$,  and $Y'$. $\theta$ is a set of cosmological parameters that parametrise the HI power spectrum.   The covariance matrix of the measured power spectrum, Cov$(XX',YY')$, is
\begin{equation}
\begin{split}
\left[\text{Cov}(XX',YY')\right]_{\ell} = & \frac{1}{(2\ell+1)f_{\text{sky}}}  \\
& \left(\hat{C_{\ell}}^{XY}\hat{C_{\ell}}^{X'Y'}+\hat{C_{\ell}}^{XY'}\hat{C_{\ell}}^{X'Y}\right)\,,
\end{split}
\label{equcov}
\end{equation}
where $f_{\rm sky}$ is the fractional sky coverage of the survey, and $\hat{C_{\ell}}$ is the measured HI power spectrum including noise such that
\begin{equation}
\hat{C_{\ell}}(z_i,z_j) = C_{\ell}^{\rm HI}(z_i,z_j) + N_{\ell}(z_i,z_j)B_{\ell}(z_i, z_j)\,,
\label{equno1f}
\end{equation}
including thermal noise only, and 
\begin{equation}
\hat{C_{\ell}}(z_i,z_j) = C_{\ell}^{\rm HI}(z_i,z_j) + \left[N_{\ell}(z_i,z_j) + F_{\ell}(z_i, z_j)\right]B_{\ell}(z_j, z_j)\,,
\label{equyes1f}
\end{equation}
including both thermal noise and $1/f$ noise. 

We will use the CPL model \citep{cp01, lin03} to parametrise the background equation of state for the dark sector  as
\begin{equation}
w(a) = w_0 + w_a(1-a)\,,
\label{w0}
\end{equation}
and therefore the set of cosmological parameters that we will use is 
\begin{equation}
  \boldsymbol{\theta} = \{\Omega_bh^2, \Omega_ch^2, w_0, w_a, h, n_s, \mathrm{log}(10^{10}A_s), b_{\rm HI}\}\,.
\label{basepar}
\end{equation}
The fiducial values of these parameters are adopted from \cite{odb+18} for consistency and comparison. The first row of Table\,\ref{tabglob} lists the fiducial values we use. The partial derivative of HI power spectrum  with respect to each cosmological parameter in Equ.\,\ref{equfish} is calculated numerically by varying each parameter with a step of  $\pm\Delta\theta$. The value of $\Delta\theta$ should not be too large,  so that it miscalculates the derivative, nor  too small,  so that it  introduces numerical noise.  We will use $\Delta\theta=0.5\%\times\theta$ and we have checked that the derivatives in this case are stable.  Each parameter is marginalised over other parameters, and their uncertainties  are calculated from the inverse of the Fisher matrix in Equ.\,\ref{equfish}.

The BAO scale measured from the HI power spectrum is sensitive to the Hubble rate $H(z)$ in the radial direction, and to the angular diameter distance $D_A(z)$ in the transverse direction, which both provide information revealing the  expansion history of the Universe. Thus we can constrain $H(z)$ and $D_A(z)$ at three redshifts by a coordinate transformation from  $\Omega_bh^2$, $w_0$ and $w_a$. The parameter transformation is performed through a transformation matrix $\mathcal{T}$ such that \citep{coe09}
\begin{equation}
[F'] = [\mathcal{T}]^T[F][\mathcal{T}]\,,
\label{equ:transmat}
\end{equation}
where $[F']$ is the new Fisher matrix after the parameter transformation, with new parameters $\boldsymbol{\theta'} = (\theta'_1, \theta'_2, \theta'_3, \dots)$, and $[F]$ is the old Fisher matrix with old parameters $\boldsymbol{\theta} = (\theta_1, \theta_2, \theta_3, \dots)$. The transformation matrix $\mathcal{T}$ is calculated through the partial derivative of the old parameters to the new parameters that $\mathcal{T}_{i, j} = \frac{\partial \theta_i}{\partial \theta'_j}$, and $\mathcal{T}^T$ is the transpose of the transformation matrix.   

The parameter sets after the transformation in these two cases are
\begin{equation}\label{hzpar}
\boldsymbol{\theta'} = \{H(z_0), H(z_1), H(z_2), \Omega_ch^2,  h, n_s, \mathrm{log}(10^{10}A_s), b_{\rm HI}\}\,,
\end{equation}
and
\begin{equation}\label{dapar}
\boldsymbol{\theta'} = \{D_A(z_0), D_A(z_1), D_A(z_2), \Omega_ch^2,  h, n_s, \mathrm{log}(10^{10}A_s), b_{\rm HI}\}\,.
\end{equation}
 Therefore, our constraints on $H(z)$ and $D_A(z)$ are limited to three redshift values through the parameter transformation method. In each case,  the parameter is marginalised over as a free parameter in each redshift bin.  We choose three particular redshift bins at $z = [0.5, 1.5, 2.5]$ for Band\,1 and $z = [0.05, 0.2, 0.4]$ for Band\,2, because they properly sample the shape and amplitude of the partial derivative  $\frac{\partial\theta_i}{\partial \theta'_j}$, encountered during the coordinate transformation with $\theta_i\in[\Omega_bh^2, w_0, w_a]$ and  $\theta'_j\in[H(z), D_A(z)]$.  We have tested other redshift bins under the same selection criteria,  which all give consistent results. 

The linear growth rate $f(z)$, characterising the RSD effect,  can be used for testing alternative theories of gravity, which alters galaxy peculiar velocities with respect to the General Relativity prediction \citep[e.g.,][]{jz08, bfs14}.  Therefore it is also useful to  project constrains on $f(z)$, and understand the impact of $1/f$ noise. We forecast constraints on $f\sigma_8(z) = f(z)\sigma_8D(z)$ with  10 equally-spaced frequency bins expanding over the full band of SKA1-MID Band\,1 and Band\,2, respectively.  We assume a piece-wise linear parametrization of $f\sigma_8(z)$ at the 10 frequency bins. We vary the amplitude of each bin with  $\Delta f\sigma_8(z)=0.5\%\times f\sigma_8(z)$, and calculate the the derivative of the HI spectrum with respect to $f\sigma_8(z)$ numerically from there.  The combination of  $f(z)\sigma_8D(z)$  takes into account the degeneracy between $f(z)$, $D(z)$ and $\sigma_8$. The parameter set in this case is
\begin{equation}
\boldsymbol{\theta'} = \{f\sigma_8(z_1), f\sigma_8(z_2), \dots, f\sigma_8(z_{10}) \}\,.
\end{equation}
Each of the 10 $f\sigma_8(z)$  is treated as an independent parameter  and marginalised over, with all other base parameters fixed.  

\subsection{\emph{Planck} prior}\label{sec:prior}
Although  SKA1-MID IM survey is promising in terms of probing the dark sector, it will find it difficult to constrain all parameters by itself. Combining datasets from other probes can break parameter degeneracies and improve precision. The high precision measurements of the CMB from \emph{Planck} already provide very tight constraints on the  base parameters of $\Lambda$CDM model. Therefore, we will include a  \emph{Planck} prior, which is obtained from \emph{Planck} 2015 TT+TE+EE likelihood data \citep{planckXI15} using \textsc{CosmoMC} \citep{lb02}, that uses the Markov Chain Monte Carlo technique to estimate the maximum-likelihood value for cosmological parameters and return their covariance. In particular, we take the output parameter covariance matrix from \textsc{CosmoMC},  and add it into our IM Fisher matrix. We assume a $w$CDM model when producing the \emph{Planck} prior, since the high redshift CMB measurement reveals little information on $w_a$.  We include, from the output covariance matrix from \textsc{CosmoMC},  only the 6 relevant parameters  for our analysis such that
\begin{equation}
\boldsymbol{\mathcal{P}}=\{\Omega_bh^2, \Omega_ch^2, n_s, \mathrm{log}(10^{10}A_s), w_0, h\}\,,
\end{equation}
 and treat all other \textsc{CosmoMC} parameters as nuisance parameters.


\section{Projected cosmological parameter constraints for thermal noise only}\label{sec:thermalonly}
In this section, we present the results of cosmological parameter constraints from the Fisher matrix analysis, where only thermal noise is  included without $1/f$ noise. We present the  results in Sect.\,\ref{sec:baseno1f}. We investigate the importance of accurate HI power spectrum calculation, in terms of the \emph{redshift-space-distortion} component (Sect.\,\ref{sec:rsd}), cross-frequency components (Sect.\,\ref{sec:crossbin}), and the total number of frequency channels (Sect.\,\ref{sec:nbin}).
\begin{table*}
\centering
\begin{tabular}{l|l|l|l|l|l|l|l|l}
\hline
\hline
Parameters & $\Omega_bh^2$ & $\Omega_ch^2$ & $w_0$ & $w_a$ & $h$ & $\log (10^{10}A_s)$ & $n_s$ & $b_{\rm HI}$ \\
&  [0.02224] & [0.1198] & [-1.00] & [0.00] & [0.6727]  & [3.096] & [0.9641] & [1.00]  \\
\hline
\emph{Planck} & $\pm0.00015$ & $\pm0.0016$  & $\pm0.45$  & $\dots$ & $\pm0.13$ & $\pm0.039$ &   $\pm0.0046$ & $\dots$ \\ \hline
Band\,1 alone  & $\pm0.0037$  & $\pm0.013$  & $\mathbf{\pm0.061}$ & $\mathbf{\pm0.23}$ & $\mathbf{\pm0.035}$ & $\pm0.11$ & $\pm0.038$ &$\mathbf{\pm0.033}$  \\
& (2.9) & (3.0) &(1.4) &(1.9) & (2.9) & (2.4) & (2.3) & (2.1) \\ \hline
Band\,2 alone & $\pm0.0039$ & $\pm0.0085$ & $\mathbf{\pm0.15}$ & $\mathbf{\pm0.79}$ & $\mathbf{\pm0.023}$ & $\pm0.30$ & $\pm0.014$ & $\mathbf{\pm0.19}$ \\
& (1.7) & (2.3) & (1.5) & (1.5) & (2.6) & (1.7) & (1.7) & (1.5) \\ \hline
Band\,1+\emph{Planck}  & $\pm0.00012$ & $\pm0.00092$ & $\mathbf{\pm0.036}$ & $\mathbf{\pm0.12}$& $\mathbf{\pm0.0095}$ & $\pm0.032$& $\pm0.0038$ & $\mathbf{\pm0.017}$ \\
 & (1.2) & (1.5) & (1.3) & (1.3) & (1.3) & (1.1) & (1.1) & (1.2) \\ \hline
Band\,2+\emph{Planck} & $\pm0.00013$ & $\pm0.0010$  & $\mathbf{\pm0.032}$ & $\mathbf{\pm0.17}$ & $\mathbf{\pm0.0031}$ & $\pm0.038$   & $\pm0.0037$ & $\mathbf{\pm0.022}$\\   
 & (1.1) & (1.4) &  (1.6) & (1.3) & (1.0) & (1.0) & (1.1) & (1.0) \\  \hline
Band\,1+Band\,2 & $\pm0.00012$ & $\pm0.00081$ &$\mathbf{\pm0.024}$ & $\mathbf{\pm0.099}$ & $\mathbf{\pm0.0024}$ & $\pm0.030$ & $\pm0.0034$ & $\mathbf{\pm0.016}$ \\
+\emph{Planck} & & & & & & & & \\ \hline
noRSD & $\pm0.0073$ & $\pm0.025$ &$\pm0.20$ & $\pm0.58$ & $\pm0.068$ & $\dots$ & $\pm0.065$ & $\dots$ \\ \hline
noRSD+\emph{Planck} &   $\pm0.00014$ & $\pm0.0013$ & $\pm0.084$ & $\pm0.25$ & $\pm0.014$ &   $\pm0.039$ & $\pm0.0043$ & $\pm0.020$ \\
\hline
\hline
\end{tabular}
\caption{The cosmological parameters (\emph{1st row}) and their projected uncertainties from \emph{Planck} likelihood (\emph{2nd row}) using \textsc{CosmoMC}, SKA Band\,1 alone (\emph{3rd row}),  SKA Band\,2 alone (\emph{4th row}),  SKA Band\,1+\emph{Planck} (\emph{5th row}), SKA Band\,2+\emph{Planck} (\emph{6th row}), SKA Band\,1+Band\,2+\emph{Planck} (\emph{7th row}), SKA Band\,1 without RSD component (\emph{8th row}), and \emph{Planck}+SKA Band\,1 without RSD component (\emph{8th row}). Square brackets in the \emph{1st row} are the assumed central values of cosmological parameters.  We include only thermal noise in the SKA IM noise calculation without $1/f$ noise.  The parameters which are significantly improved  by the inclusion of  SKA IM  are highlighted in  \emph{bold}.  The numbers in  brackets give the degradation factor of the corresponding  parameter constraint after excluding the cross-correlation signal between frequency channels in the HI power spectrum calculation. }
\label{tabglob}
\end{table*}

\subsection{SKA1-MID projections}\label{sec:baseno1f}
In this section, we present projected  constraints on the 8 cosmological parameters (Equ.\,\ref{basepar}) using the Fisher matrix method described in Sect.\,\ref{sec:fm} for the case of thermal noise only. We do this for  both the SKA1-MID Band\,1 and Band\,2 surveys, operating in single dish mode, with the survey parameters given in Table\,\ref{skainput}. The projected uncertainties on the  cosmological parameters are presented in Table\,\ref{tabglob} along with the \emph{Planck} priors used in our analysis (Sect.\,\ref{sec:prior}). 

SKA1-MID results are presented in  the \emph{3rd} and \emph{4th} rows of Table\,\ref{tabglob}. The constraints on $w_0$ and $h$ are significantly improved compared to \emph{Planck}. In particular, SKA1-MID Band\,1 alone has the potential to constrain $w_0$ with $\approx 6\%$ accuracy, and $h$ with $\approx 3\%$ accuracy. This is because  the HI power spectrum measured by SKA IM over a range of redshifts breaks the angular diameter distance degeneracy inherent in CMB measurements, allowing simultaneous measurement of  $w_0$, $w_a$ and $h$. In addition, the amplitude of the HI spectrum is sensitive to $b_{\rm HI}$, allowing a $\approx 3\%$ constraint using Band\,1 alone, in line with that reported in \cite{skaredbook18}.  

\begin{figure}
\centering
\includegraphics[width=0.95\hsize]{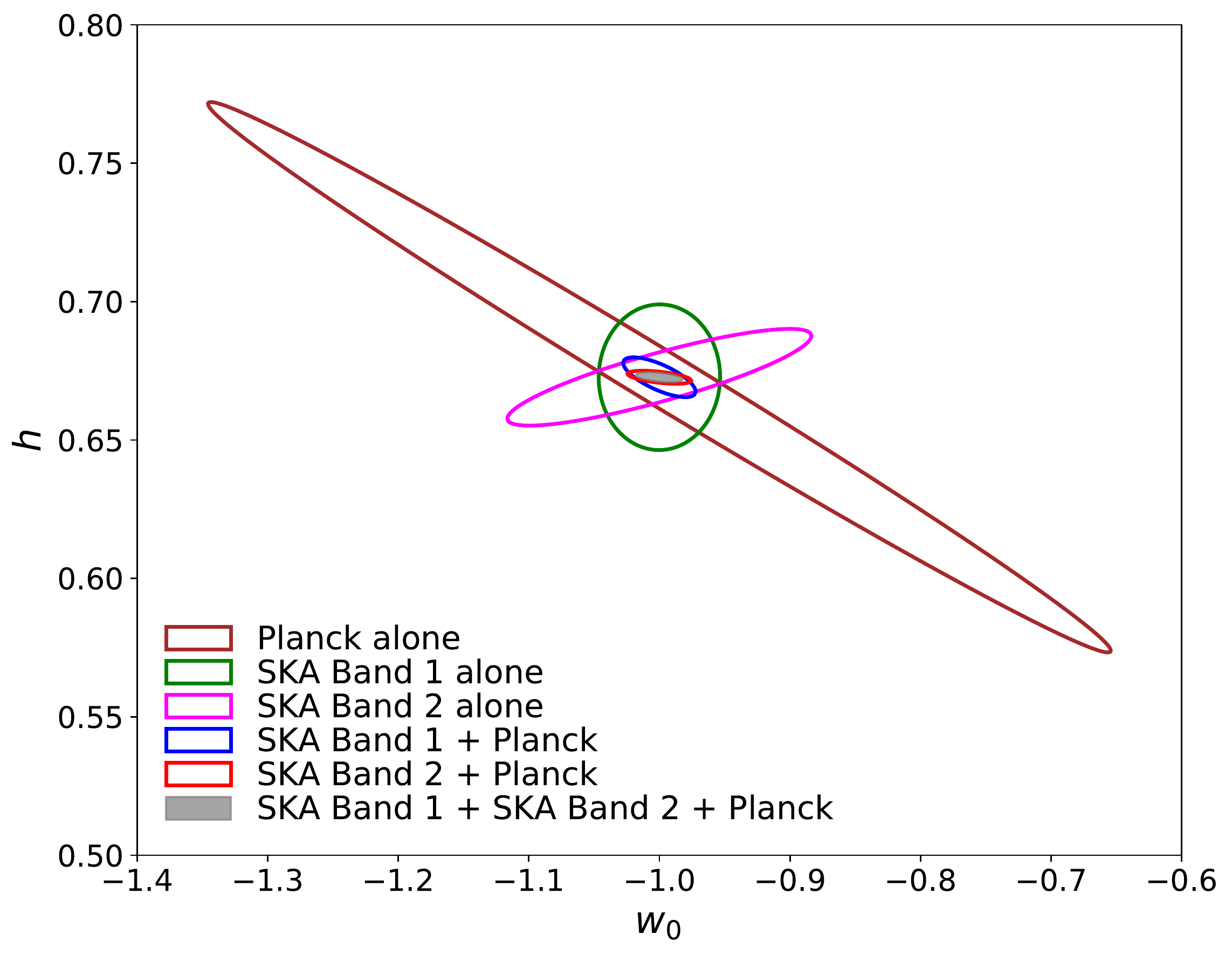}
\caption{The joint constraints on $w_0-h$ for \emph{Planck} alone (\emph{brown}), SKA1-MID Band\,1 alone (\emph{green}), Band\,2 alone (\emph{magenta}), Band\,1+\emph{Planck} (\emph{blue}), Band\,2+\emph{Planck} (\emph{red}), and Band\,1+Band\,2+\emph{Planck} (\emph{filled grey}). All other cosmological parameters have been marginalised. We include only thermal noise for SKA IM. The degeneracy between $w_0$ and $h$ is broken by the addition of $\emph{Planck}$ and SKA IM. }
\label{fig:hw}
\end{figure}

Comparing Band\,1 with Band\,2, we find that a tighter constraint would be obtained  from Band\,1 on $w_0$, $w_a$ and $b_{\rm HI}$, but from Band\,2 on $h$. This is because a survey with Band\,2 will be more sensitive to lower redshifts, yielding tighter constraints on $h$ than with Band\,1, whose wider and higher redshift range facilitate measurements of $w_0$ and $w_a$. In Fig.\,\ref{fig:hw} we show  the joint constraint on $w_0-h$ for the various surveys presented in Table\,\ref{tabglob}, illustrating a clear degeneracy between the two parameters from the \emph{Planck} measurement. In contrast, tighter constraints on both parameters are projected from measurements using SKA IM surveys with different degenerate lines. In particular, SKA Band\,1 alone is powerful in terms of constraining $w_0$, while Band\,2 alone is better at constraining $h$.

Note that IM observations are  sensitive to $\Omega_bh^2$, $\Omega_ch^2$, $\log A_s$ and $n_s$, but constraints are orders of magnitude weaker than those from \emph{Planck}. Therefore, we add the \emph{Planck} prior to the SKA Fisher matrices and the results are presented in  the \emph{5th}, \emph{6th} and \emph{7th} rows of Table\,\ref{tabglob} for Band\,1+\emph{Planck} ,  Band\,2+\emph{Planck} and Band\,1+Band\,2+\emph{Planck}, respectively. We find that this constrains the base $\Lambda$CDM parameters and further breaks  degeneracies between $h$, $w_0$, $w_a$.  The projected constraint on $w_0$ is further improved to $\approx 3\%$ for SKA +\emph{Planck}. This can be further illustrated from Fig.\,\ref{fig:hw} where the addition of \emph{Planck} and SKA IM measurements breaks the angular diameter distance degeneracy,  leading to significantly smaller contours on the joint $w_0-h$ plane.  

Similar results have been reported previously in \cite{odb+18} where a $4\%$ and $2\%$ constraint was obtained from Band\,1+\emph{Planck} and Band\,2+\emph{Planck} respectively, using a maximum likelihood analysis for  the $w$CDM model from simulated maps, and for a  slightly different cosmological parameter set to ours. In addition, the projected  constraint on the HI bias is improved to $\approx 2\%$ for both bands after adding the \emph{Planck} prior, because it breaks the degeneracy between $\log A_s$ and $b_{\rm HI}$.  

Our results show that under the assumption of completely Gaussian white noise without systematics or foreground contamination, an intensity mapping survey with SKA1-MID combined with \emph{Planck} has the potential to tightly constrain cosmological parameters under the CPL model, nourishing the Dark Energy study. 

\begin{figure*}
\begin{center}
\includegraphics[width = 0.48\hsize]{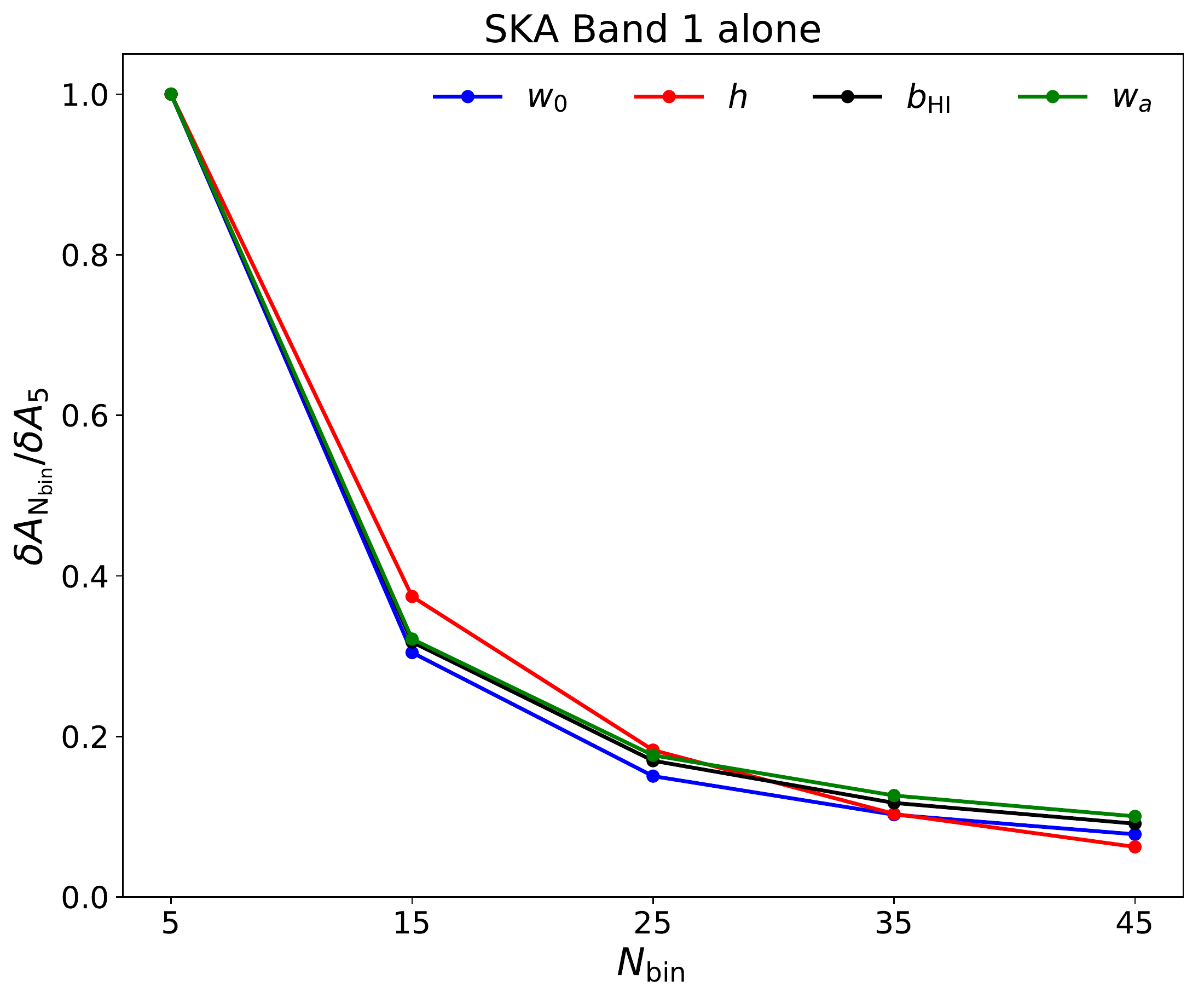}
\includegraphics[width = 0.48\hsize]{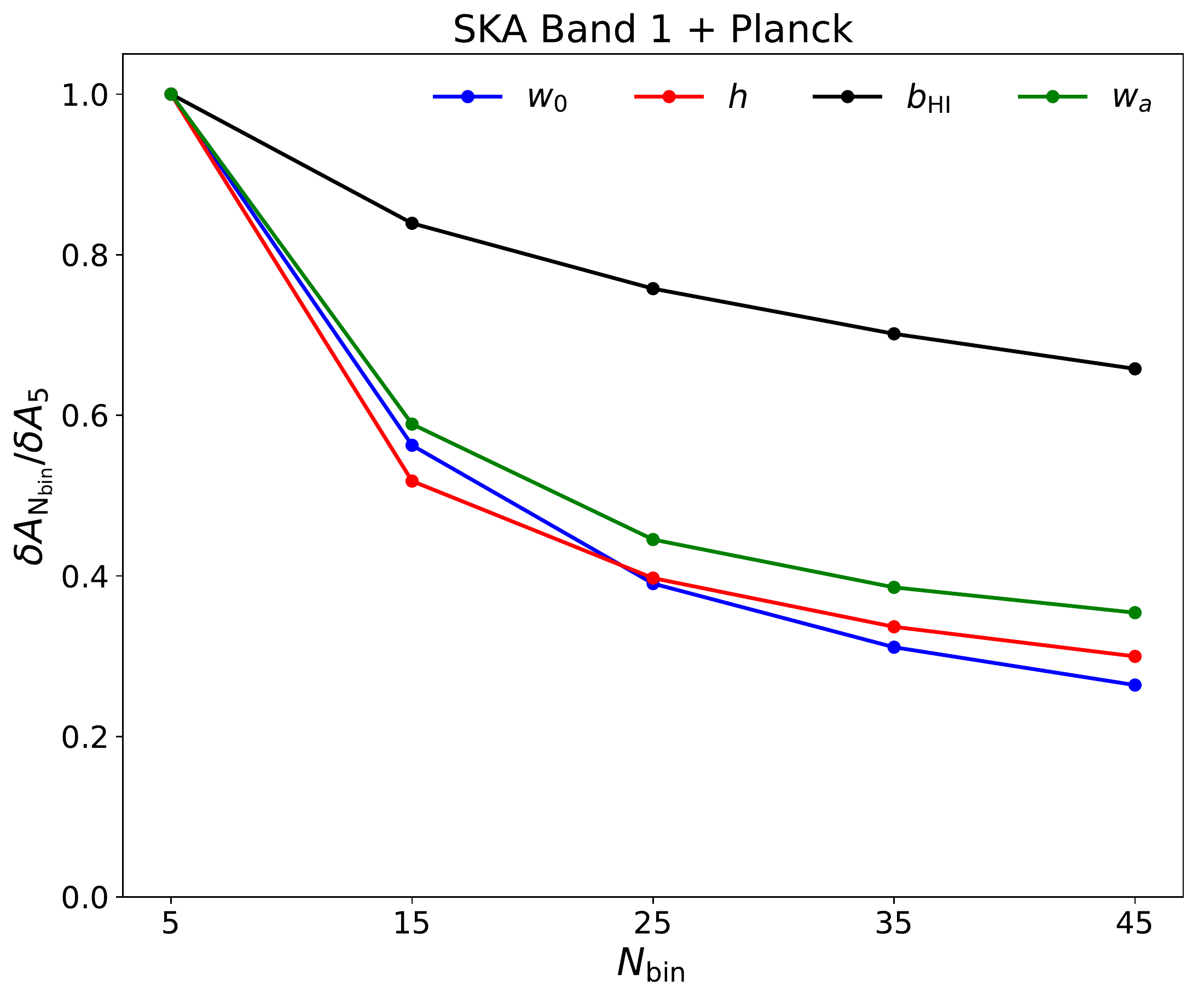}
\end{center}
\caption{Cosmological parameter constraints relative to those with $N_{\rm bin} = 5$ as a function of the number of frequency channels for  SKA Band\,1 (\emph{left}) and \emph{Planck}+Band\,1 (\emph{right}).  We vary $N_{\rm bin}$ in the range of [5, 15, 25, 35, 45]. The uncertainties decrease with increased $N_{\rm bin}$ until $N_{\rm bin} = 25$, and flatten beyond that, suggesting that the choice of $N_{\rm bin} = 35$ used in this paper does not lose significant information.} 
\label{fignbin}
\end{figure*}

\subsection{Importance of redshift-space-distortions} \label{sec:rsd}
From Equ.\,\ref{eq:Cell2}, we see that the RSD component, encoded in the  $f(z)$ term, is the key to  breaking the complete degeneracy between $A_s$ and $b_{\rm HI}$. We would expect no constraint on $A_s$ or $b_{\rm HI}$ in the absence of the RSD component in an IM only survey.  In order to verify this, we artificially set the $f(z)$ term to be zero and recalculate parameter constraints from SKA1-MID Band\,1. We only present results for Band\,1, but find the same results for Band\,2, and our main  conclusions are independent of the exact band analysed. Therefore, throughout the paper, unless otherwise stated, we will present results for Band\,1 only.  The results without the RSD component are given  in the last two rows of Table\,\ref{tabglob}, with infinite uncertainties on $A_s$  or $b_{\rm HI}$.  Adding the \emph{Planck} prior provides a constraint on $A_s$ and thus enables  a measurement on  $b_{\rm HI}$.  

Our results confirm that the RSD component is essential to obtain a constraint on $A_s$ and $b_{\rm HI}$. Therefore, the small uncertainties on  $A_s$  and $b_{\rm HI}$ given by  \cite{odb+18} are a result of the inclusion of a  \emph{Planck}  prior,  since the RSD component was neglected  in their HI power spectrum calculation.

\subsection{Cross-frequency contribution}\label{sec:crossbin}
In order to understand the importance of  the cross-correlation HI signal between frequency channels, we have performed an analysis only including  the auto-correlation signal for each frequency channel in the HI power spectrum. We expect that the parameter uncertainties will degrade, compared with those obtained using the full power spectrum calculation, due to the loss of information.

In Table\,\ref{tabglob}, we present (in brackets) the degradation factor of each parameter below its uncertainty obtained using the full power spectrum calculation. The degradation factor for each cosmological parameter is defined as the ratio of its uncertainty without the cross-correlation signal to that with the full power spectrum calculation. We see that the cross-correlation signal is more important to Band\,1 than in Band\,2 since Band\,1 has more frequency bins  and thus suffers more loss when the cross-correlation signal is excluded. In the case where the \emph{Planck} prior is added, the exclusion of the  cross-correlation signal makes less difference. These results are consistent with the expectation, and confirm that one shall adopt the full HI power spectrum calculation  wherever possible for a more accurate forecast.  

\begin{figure*}
\begin{center}
\includegraphics[width = 0.48\hsize]{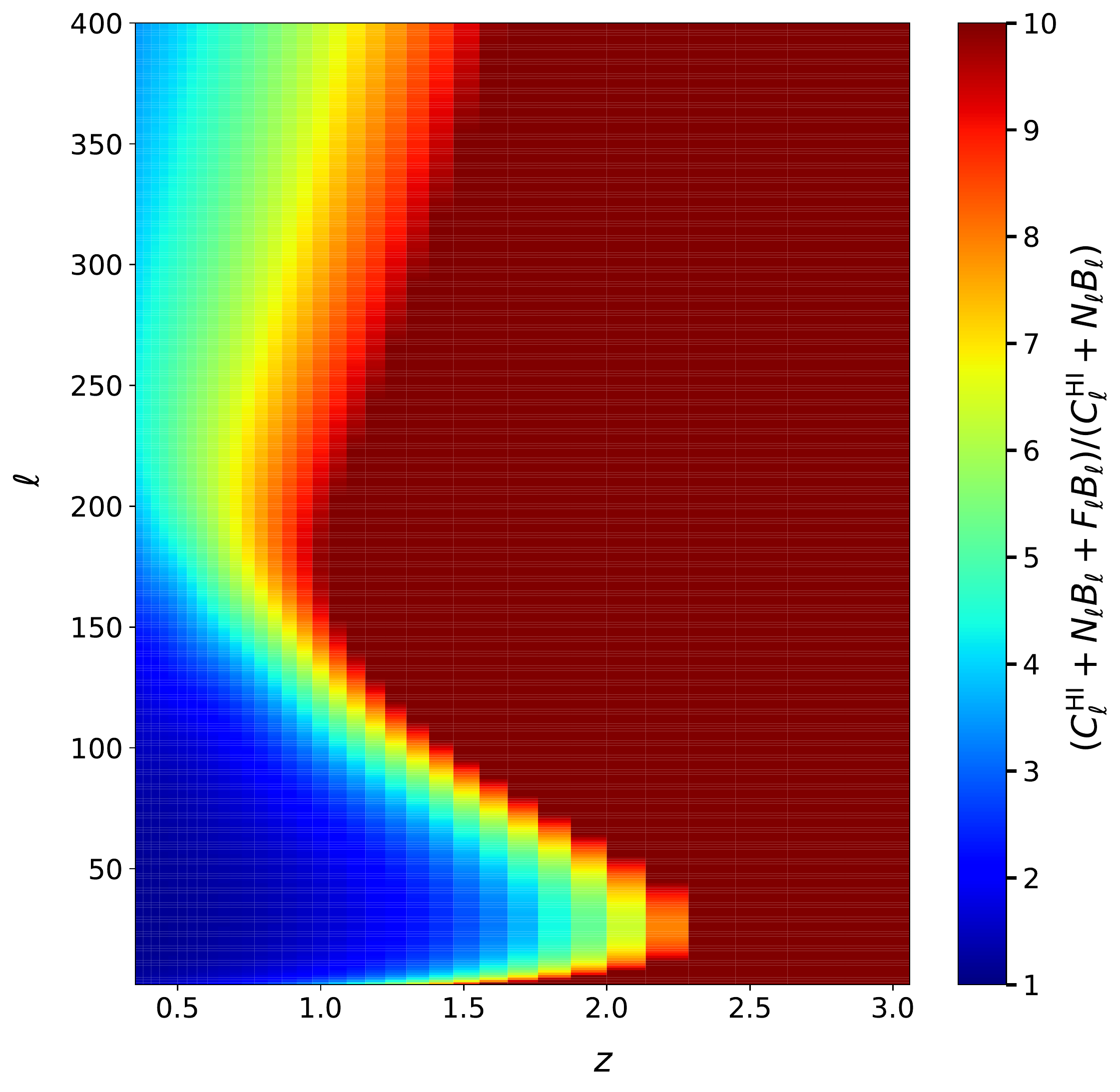}
\includegraphics[width = 0.48\hsize]{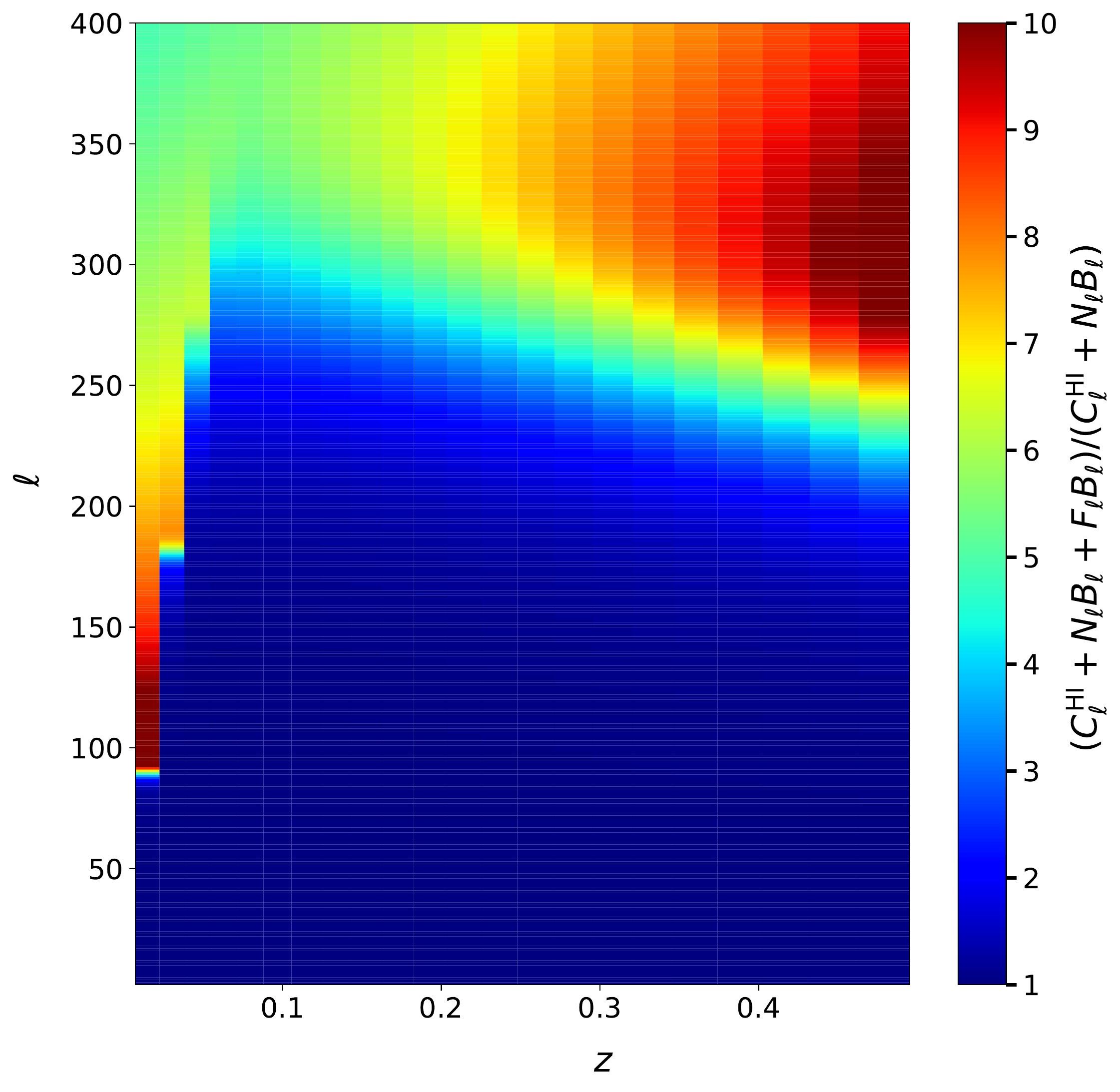}
\end{center}
\caption{The degradation factor  $\delta = (C_{\ell}^{HI}+N_{\ell}B_{\ell} + F_{\ell}B_{\ell})/(C_{\ell}^{HI}+N_{\ell}B_{\ell} )$ of SKA Band\,1 (\emph{left}) and Band\,2 (\emph{right}) as a function of redshift and multipole.  The baseline $1/f$ noise has the parameter set of [$\beta = 0.5,\;v_{\rm t} = 1\,\mathrm{deg\,s^{-1}},\;f_k = 1\,\mathrm{Hz},\;\alpha = 1$]. Higher redshifts beyond beam scales are most affected by the $1/f$ noise. Here the color scale is saturated to a maximum $\delta = 10$ beyond which little constraining power can be obtained.}
\label{figcldia}
\end{figure*}
\subsection{Impact of number of frequency channels} \label{sec:nbin}
We have chosen a specific number of frequency bins, $N_{\rm bin}$, for our analysis: 35 in the case of Band\,1 and 22 for Band\,2. In principle, we could have used any value but as $N_{\rm bin}$ decreases, there is less information from the \emph{line-of-sight} component of the power spectrum (e.g., information on the RSD component as discussed in Sect.\,\ref{sec:rsd}).  However, as $N_{\rm bin}$ increases, the amount of information gained will reduce dramatically beyond some critical value. This is because the system noise increases $\propto\frac{1}{\sqrt{\delta\nu}}$ and thus $\propto\sqrt{N_{\rm bin}}$. Besides, computational requirements (the calculations of spectrum scale $\propto N^2_{\rm bin}$) prefer one to choose the lowest possible value.  In this section we will investigate the constraints on the cosmological parameters as a function of $N_{\rm bin}$. 

We will vary $N_{\rm bin}$ in the range of [5, 15, 25, 35, 45] keeping the total bandwidth fixed at 700\,MHz for SKA1-MID Band\,1. We would expect similar behavior for Band\,2.  The channel width in each case is simply the total bandwidth divided by $N_{\rm bin}$.   The parameter uncertainties as a function of $N_{\rm bin}$ are shown in Fig.\,\ref{fignbin}, relative to  their uncertainty $\delta A_5$  for $N_{\rm bin} = 5$. Since we have shown in Table\,\ref{tabglob} that  $w_0$, $w_a$,  $h$ and $b_{\rm HI}$ are those parameters most strongly constrained by the IM data, hereafter we will only present results for these parameters. The left panel in Fig.\,\ref{fignbin} is  from Band\,1 alone, and the right panel is from \emph{Planck}+Band\,1. In both cases, we observe significant improvement on parameter constraints from $N_{\rm bin}=5$ to $N_{\rm bin} = 25$, but little additional information beyond $N_{\rm bin} = 25$ because of the increases noise.  The parameter constraints in the case of \emph{Planck}+Band\,1 are less strongly impacted by varying $N_{\rm bin}$ than Band\,1 alone.


\section{Impact of $1/f$ noise} \label{sec:with1f}
In this section, we investigate the impact of $1/f$ noise on intensity mapping, adopting the semi-empirical $1/f$ noise model introduced in Sect.\,\ref{sec:1fintro} based on \cite{hdb+18}. We quantify the consequent degradation caused by $1/f$ noise on  the power spectrum (Sect.\,\ref{sec:ps}), the impact of different $1/f$ noise parameters (Sect.\,\ref{sec:basepar}), and  the consequent degradation on the constraints of $w_0-w_a$ (Sect.\,\ref{sec:w0-wa}), $H(z)$ and $D_A(z)$ (Sect.\,\ref{sec:hz}),  and $f(z)$ (Sect.\,\ref{sec:fz}).

\subsection{Power spectrum degradation}\label{sec:ps}
In order to understand the impact of $1/f$ noise on power spectrum measurements, we define a   degradation factor, $ \delta_{\ell}(z_i, z_j)$, as the ratio of the total power spectrum with $1/f$ noise (Equ.\,\ref{equyes1f}) to that without $1/f$ noise (Equ.\,\ref{equno1f}), given by 
\begin{equation}
 \delta_{\ell} (z_i, z_j) = \frac{ C^{\rm HI}_{\ell}(z_i,z_j) + [N_{\ell}(z_i, z_j)+ F_{\ell}(z_i,z_j)]B_{\ell}(z_i, z_j)}{C^{\rm HI}_{\ell}(z_i,z_j) + N_{\ell}(z_i,z_j)B_{\ell}(z_i, z_j)}\,.
\label{sn1f}
\end{equation}
We will  only compute $\delta_{\ell}(z_i, z_j)$ for  auto-correlation frequency channels, i.e., $i=j$, since  $i \neq j$ has  $N_{\ell}(z_i, z_j) = 0$ by assumption, and a possible $C^{\rm HI}_{\ell}(z_i,z_j) = 0$ that results in an infinite $\delta_{\ell} (z_i, z_j)$. 

 The power spectrum degradation factor $\delta_{\ell} $ is plotted as a function of redshift and angular multipole in Fig.\,\ref{figcldia} for  SKA1-MID Band\,1 (\emph{left}) and Band\,2 (\emph{right}) respectively. We adopt the baseline $1/f$ noise setup at [$\beta = 0.5$, $\alpha = 1$, $f_{\rm knee} = 1\,\mathrm{Hz}$, $v_{\rm t} = 1\,\mathrm{deg\,s}^{-1}$].  For both bands, it can be seen that the degradation factor has large variations in both redshift and angular scale. Typically, it increases towards high redshift, because the HI signal is weaker at higher redshift, and thus more affected by  $1/f$ noise. This can be confirmed from Fig.\,\ref{fig:cov1f}, where the HI signal at $z = 0.5$ dominates up to the beam scale, but is completely dominated by the $1/f$ noise at $z = 2$.  Vertically, the degradation factor increases towards the beam scale ($\ell\sim200$), where the effect of   $1/f$ noise increases exponentially.
 
It is worth noting from Fig.\,\ref{figcldia} that  most of the constraining power comes from  a ``window'' at low redshift ($z\lesssim1$) below beam scales ($\ell\lesssim100$), at the presence of $1/f$ noise. The exact size of this ``window'' and its degradation factor may  vary with the $1/f$ noise level, observing frequency and is subject to the specific scan strategy and component separation analysis assumed. For SKA1-MID Band\,1,  the power spectrum measurement is  degraded by a  factor of $\delta\approx2$ at $z\lesssim1.5$ and $\ell\lesssim100$. SKA1-MID Band\,2 has a smaller degradation factor of $\delta\approx1.3$ at $z<0.5$ and $\ell\lesssim200$, due to observing at a lower redshift.

Our results show that $1/f$ noise can significantly degrade power spectrum detection, especially at high redshift. The large variation of the degradation in redshift is a big challenge to the measurement of redshift-dependent quantities, such as the growth rate $f(z)$. Therefore, one can no longer neglect $1/f$ noise for intensity mapping experiment.

\begin{figure*}
\begin{center}
\includegraphics[width = 0.5\hsize]{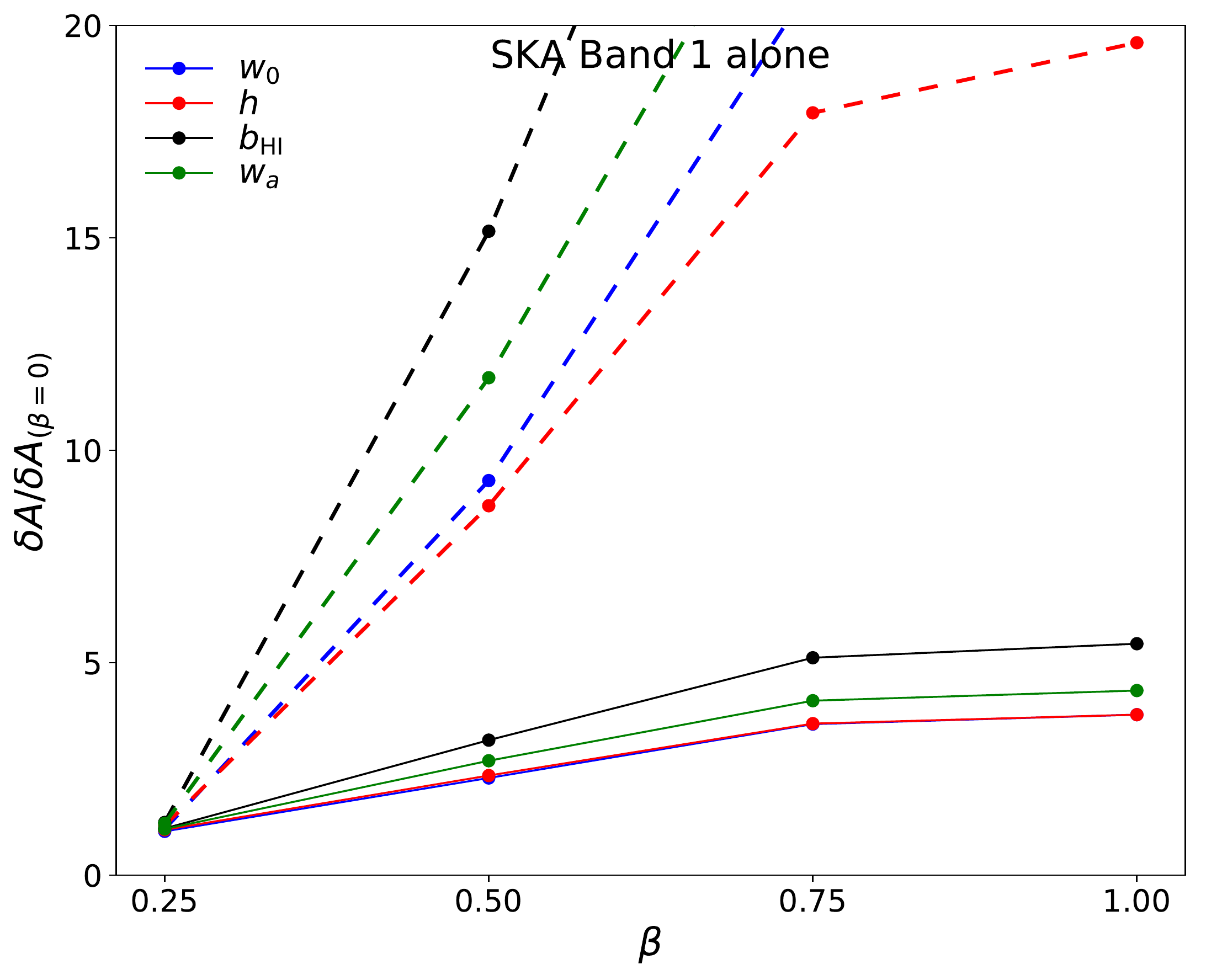}
\includegraphics[width = 0.49\hsize]{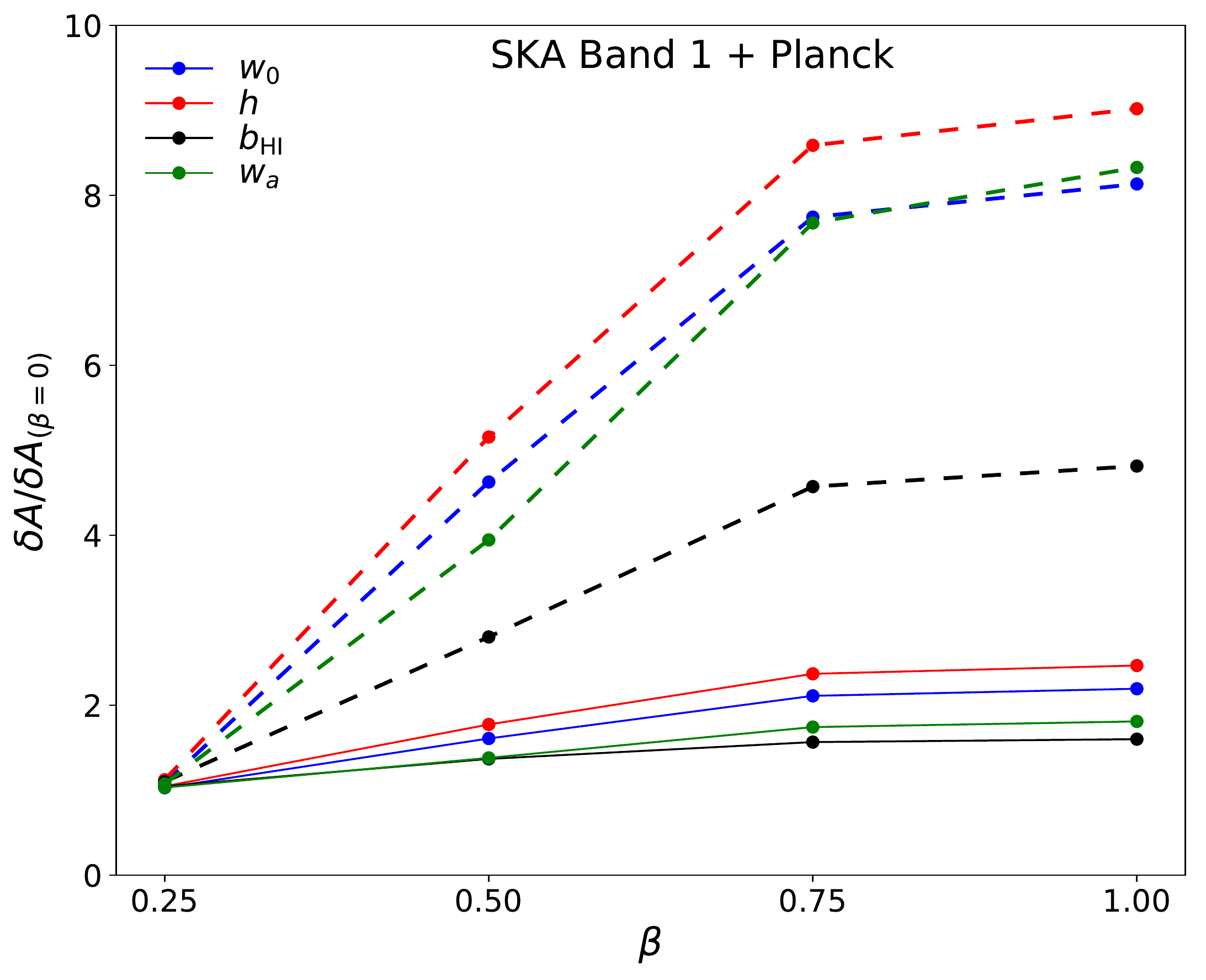}
\includegraphics[width = 0.5\hsize]{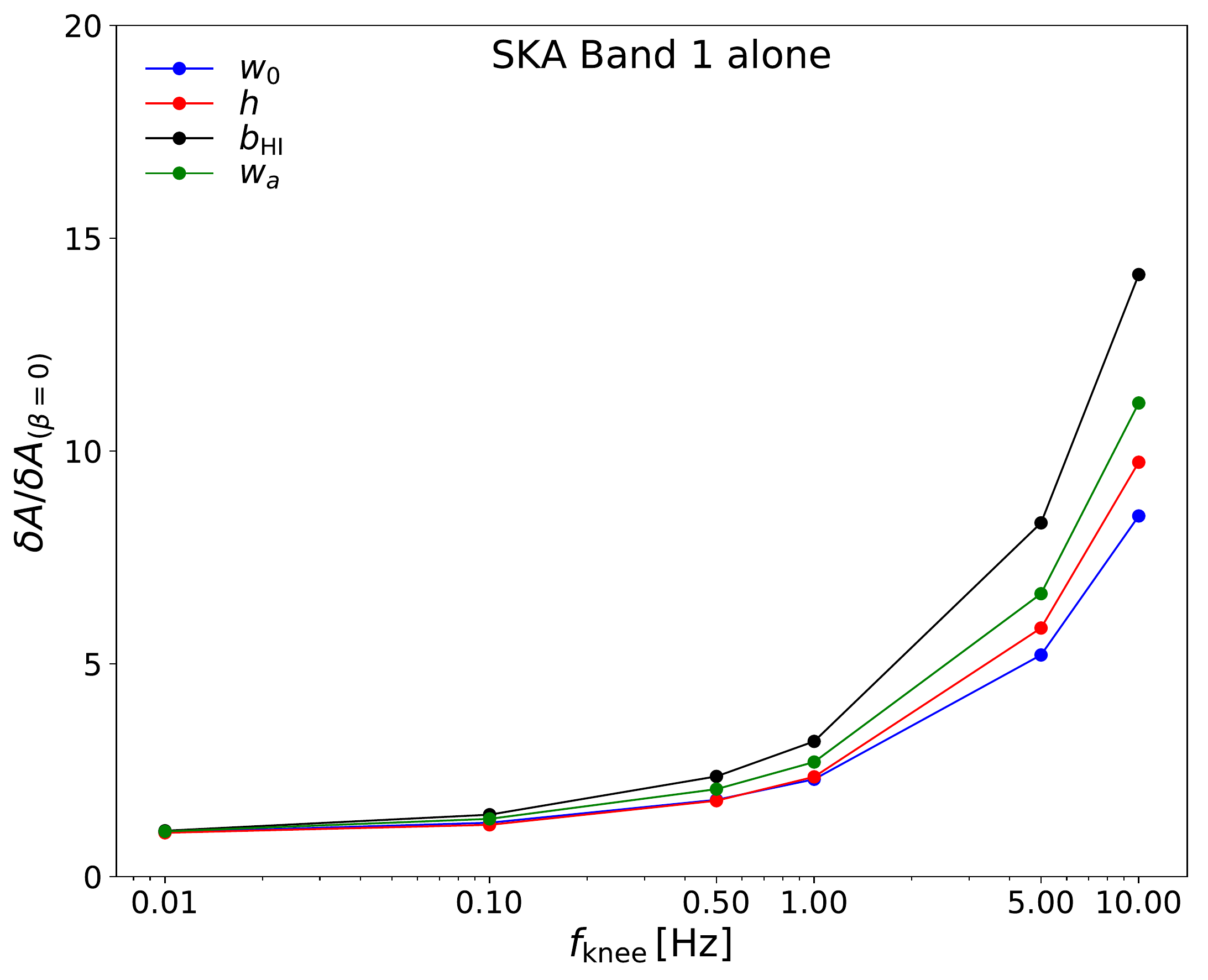}
\includegraphics[width = 0.49\hsize]{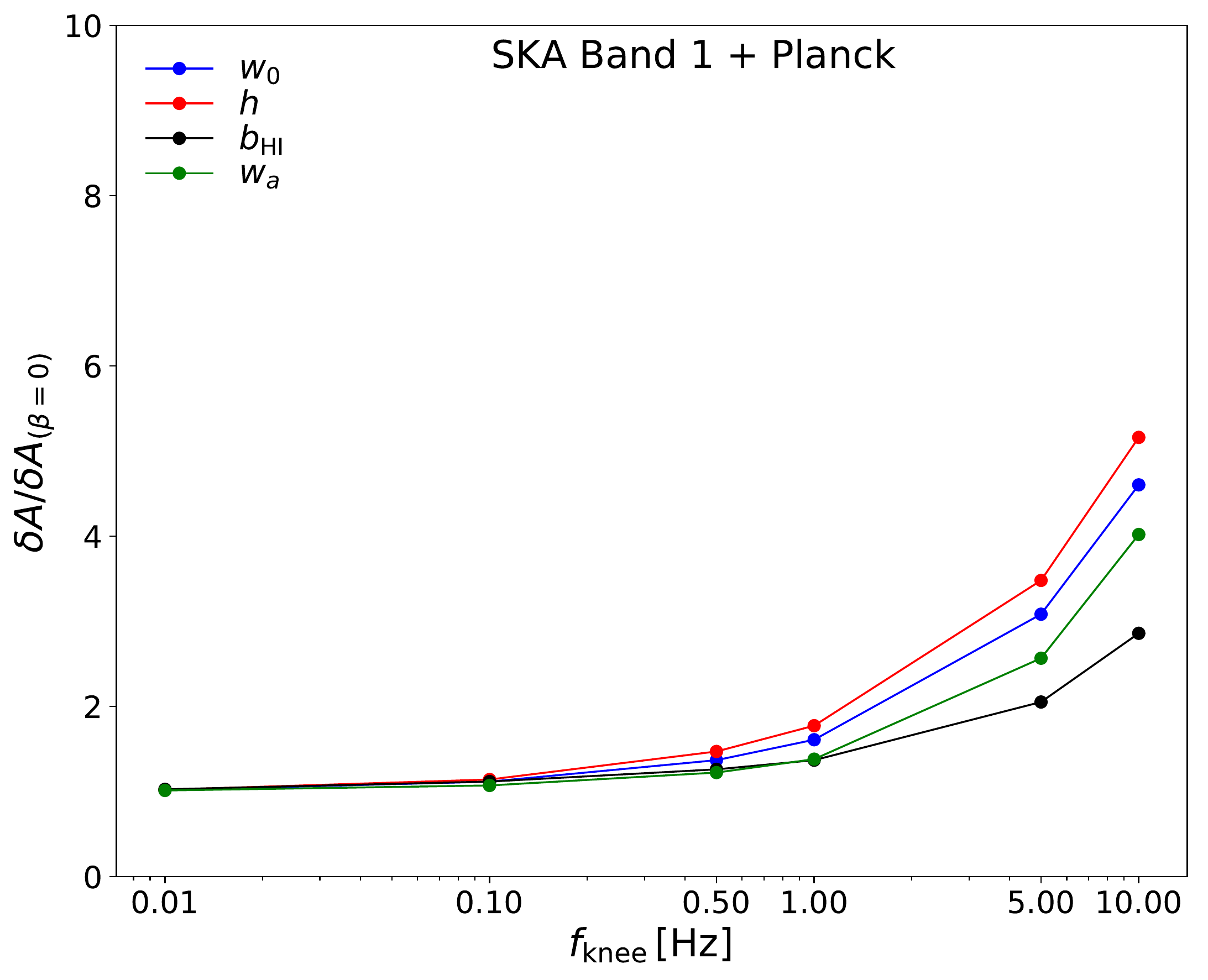}
\includegraphics[width = 0.5\hsize]{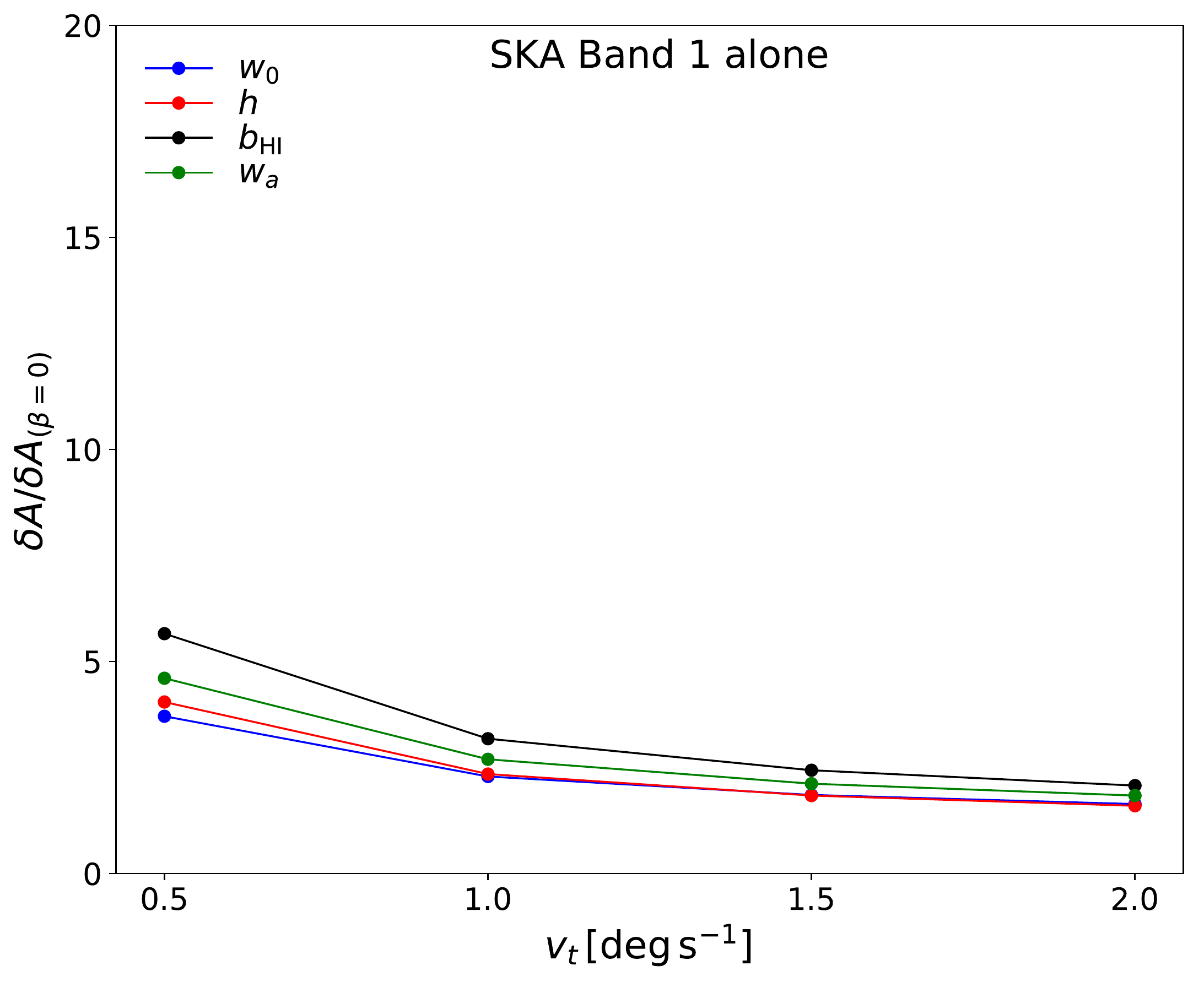}
\includegraphics[width = 0.49\hsize]{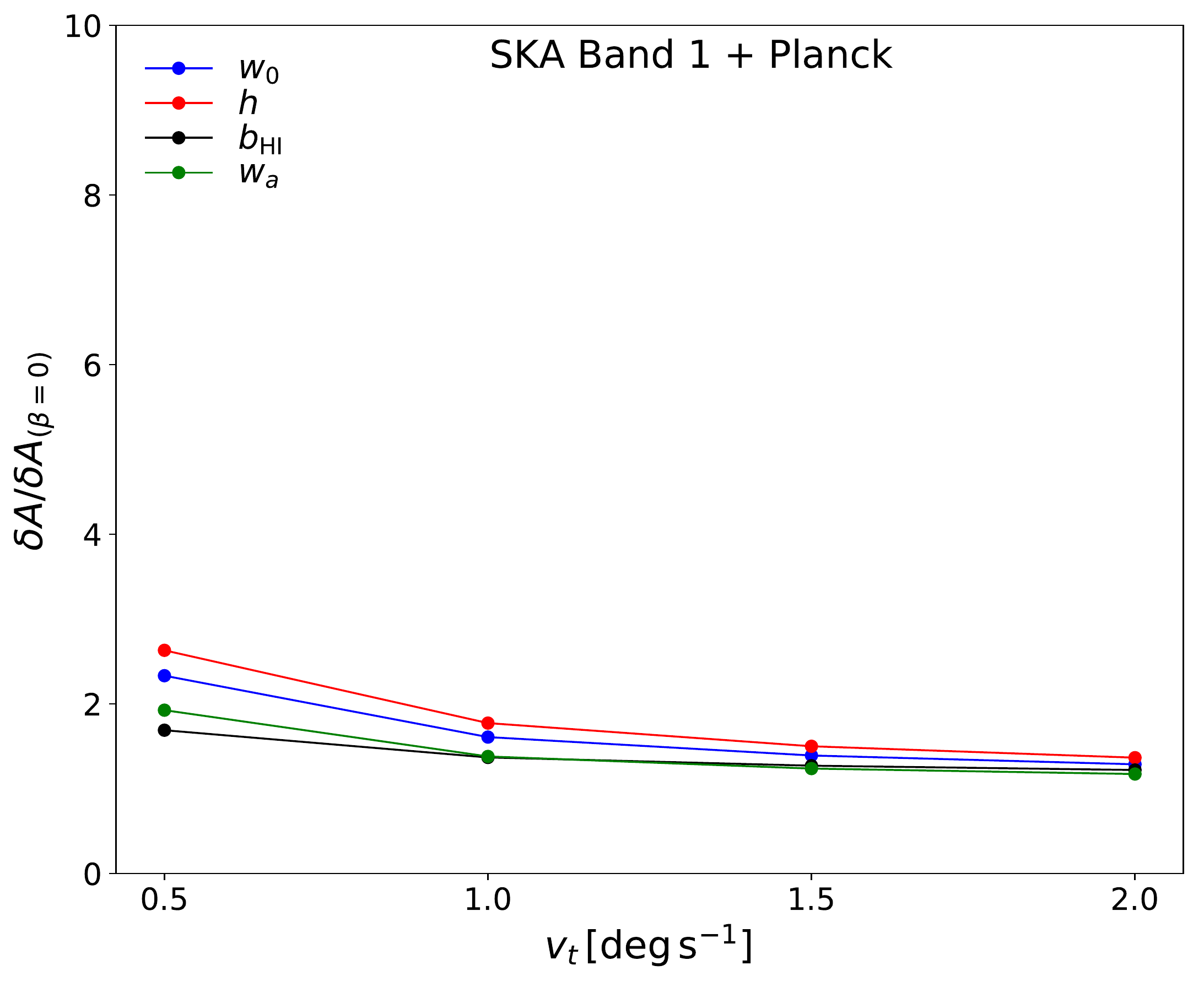}
\end{center}
\caption{The uncertainties on cosmological parameters relative to that obtained with effectively no $1/f$ noise ($\beta = 0$) as a function of the $1/f$ noise correlation index $\beta$ (\emph{top}), the knee frequency $f_{\rm knee}$ (\emph{middle}), and  the  telescope slew speed $v_{\rm t}$ (\emph{bottom}). The \emph{left} panels are obtained from SKA1-MID Band\,1 alone,  and the \emph{right} panels are from Band\,1+\emph{Planck}. In each case, the other $1/f$ noise parameters are set to the default baseline values, apart from the one that is varied. We investigate the impact of the spectral slope $\alpha$  in the \emph{top} two panels with \emph{solid} curves at $\alpha = 1$ and \emph{dashed} curves at $\alpha = 2$. The uncertainties increase towards higher values of $\alpha$, $\beta$ and $f_{\rm knee}$, but lower values of $v_{\rm t}$.  Note that we use the same \emph{y}-axis range vertically to emphasis the biggest impact from $\alpha$ and $f_{\rm knee}$. }
\label{figbeta}
\end{figure*}

\begin{figure*}
\begin{center}
\includegraphics[width = 0.48\hsize]{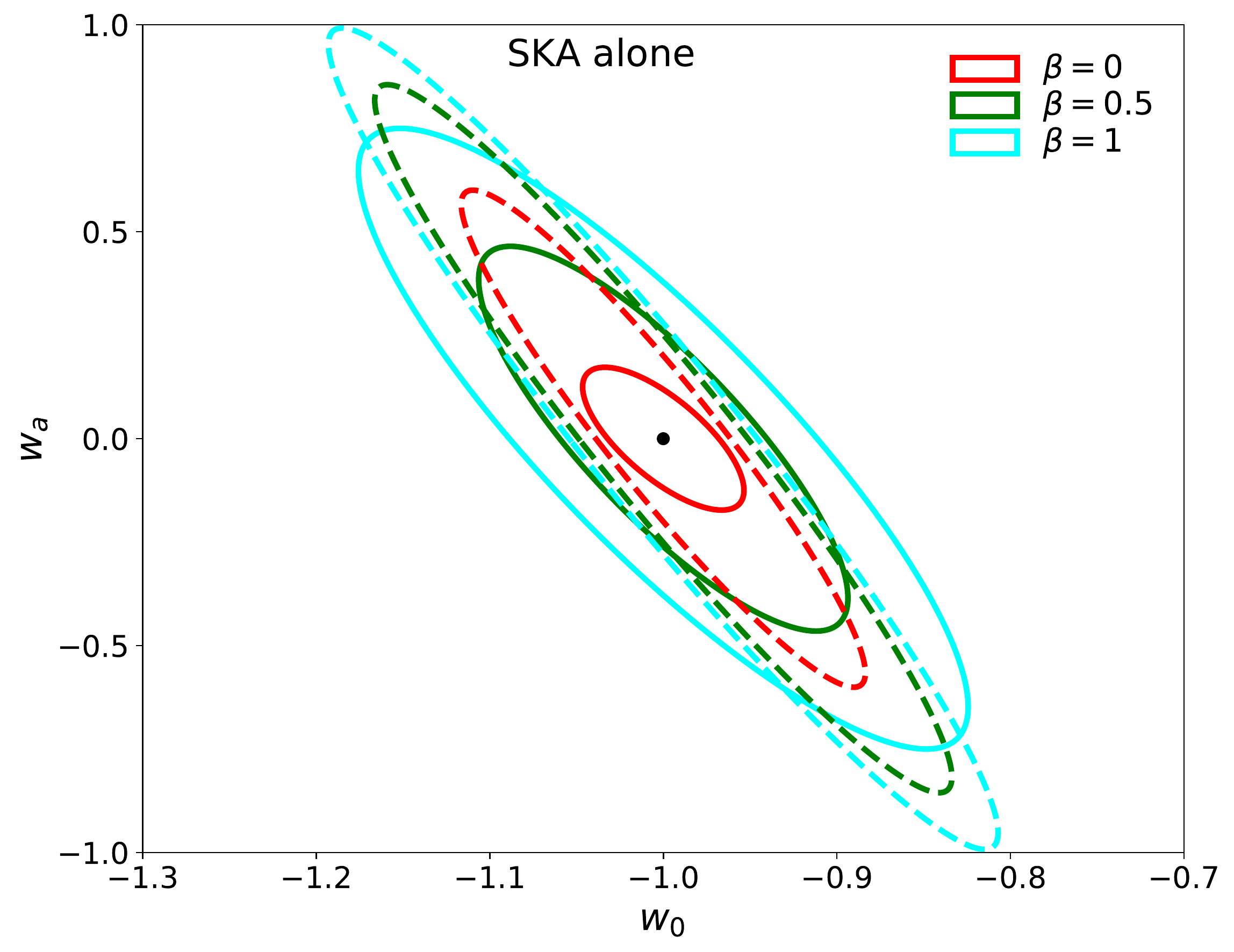}
\includegraphics[width = 0.48\hsize]{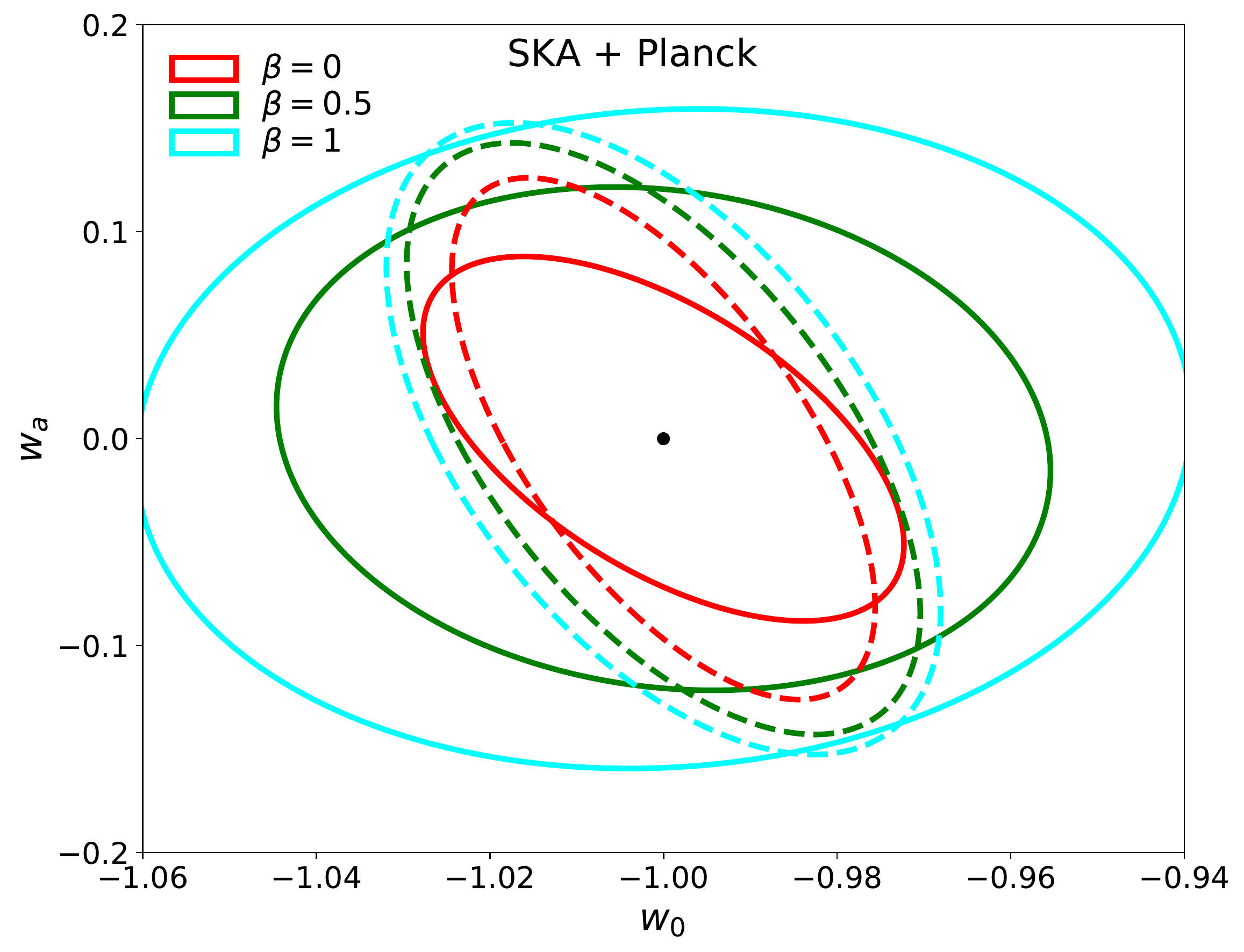}
\end{center}
\caption{The joint constraints on $w_0-w_a$ for SKA alone (\emph{left}) and SKA+\emph{Planck} (\emph{right}). All other cosmological parameters have been marginalised. In each panel, we plot the $1\,\sigma$  contours for three cases: no $1/f$ noise ($\beta = 0$); $\beta = 0.5$; $\beta = 1$. The other $1/f$ noise parameters are fixed at the baseline values of $v_{\rm t} = 1\,\mathrm{deg\,s^{-1}}$, $f_k = 1\,\mathrm{Hz}$, and $\alpha = 1$. The \emph{solid} contours are results for  Band\,1 and \emph{dashed} contours are for Band\,2. }
\label{figw0wa}
\end{figure*}
\subsection{Impact of $1/f$ noise parameters} \label{sec:basepar}
In order to understand the impact of $1/f$ noise on cosmological parameters, we vary the $1/f$ noise spectral index $\alpha$, correlation index $\beta$, knee frequency $f_{\rm knee}$, and telescope slew speed $v_{\rm t}$ respectively. We calculate the ratio of the  uncertainties on cosmological parameters relative to the case with $\beta =0$, which is the totally correlated $1/f$ noise that can be completely removed by component separation \citep{hdb+18} assuming perfect calibration and no additional systematic errors. 

The \emph{top} two panels in Fig.\,\ref{figbeta} present the results for various $\beta$ values for  SKA1-MID Band\,1 alone (\emph{left}) and Band\,1 + \emph{Planck} (\emph{right}).   The \emph{solid} curves are the results with $\alpha = 1$, and the \emph{dashed} curves are those with $\alpha = 2$, with  $f_{\rm knee}$ and $v_{\rm t}$  fixed at the baseline values of $1\,\mathrm{Hz}$ and $1\,\mathrm{deg\,s^{-1}}$ respectively. It can be seen that all cases in both panels have increased uncertainties towards  a larger value of $\beta$. By comparing $\alpha = 1$ (\emph{solid}) with $\alpha = 2$ (\emph{dashed}), the spectral index $\alpha$ increases uncertainties  significantly.  By comparing Band\,1 alone (\emph{left}) with Band\,1 + \emph{Planck} (\emph{right}),   the \emph{Planck} prior compensates for some degradation  by constraining other base parameters and breaking degeneracies.

In the middle two panels, we investigate the impact of  varying $f_{\rm knee}$, fixing other parameters at the baseline values of  $\beta = 0.5$, $v_{\rm t} = 1\,\mathrm{deg\,s^{-1}}$, and $\alpha = 1$. It can be seen that the fractional uncertainties  increase significantly as $f_{\rm knee}$ increases. In order to have a negligible  impact on cosmological parameter constraints, we deduce that one requires a  $f_{\rm knee}<0.1$\,Hz. The results from Band\,1 + \emph{Planck} (\emph{right}) are less affected by $1/f$ noise than those from Band\,1 alone (\emph{left}).

Finally we study the impact of telescope slew speed  by varying $v_{\rm t}$ in the bottom panels,  fixing other parameters at the baseline values of  $\beta = 0.5$, $f_{\rm knee} = 1\,\mathrm{Hz}$, and $\alpha = 1$.  The fractional uncertainties in this case decrease with increased slew speed as one would expect. A high  slew speed of $2\,\mathrm{deg\,s^{-1}}$ is desired if possible, in order to reduce the impact of $1/f$ noise. Again, the addition of \emph{Planck} prior mitigates part of the degradation from $1/f$ noise. 

In consistence with \cite{hdb+18}, our results show that a minimised spectral index ($\alpha << 2$) is critical, with a potential degradation by a factor of $\gtrsim10$ otherwise. A low knee frequency ($f_{\rm knee}<<0.5$\,Hz), a higher telescope slew speed, and a correlation in frequency channels are desired to diminish the effect of $1/f$ noise, although this is subject to the specific assumptions from the adopted model.  It is worth noting that even with the \emph{Planck} prior, the degradation on  $h$, $w_0$, $w_a$ and $b_{\rm HI}$ cannot be completely mitigated since they are primarily constrained by IM data, which emphases the importance of controlling $1/f$ noise for IM experiments.

\begin{figure*}
\begin{center}
\includegraphics[width = 0.48\hsize]{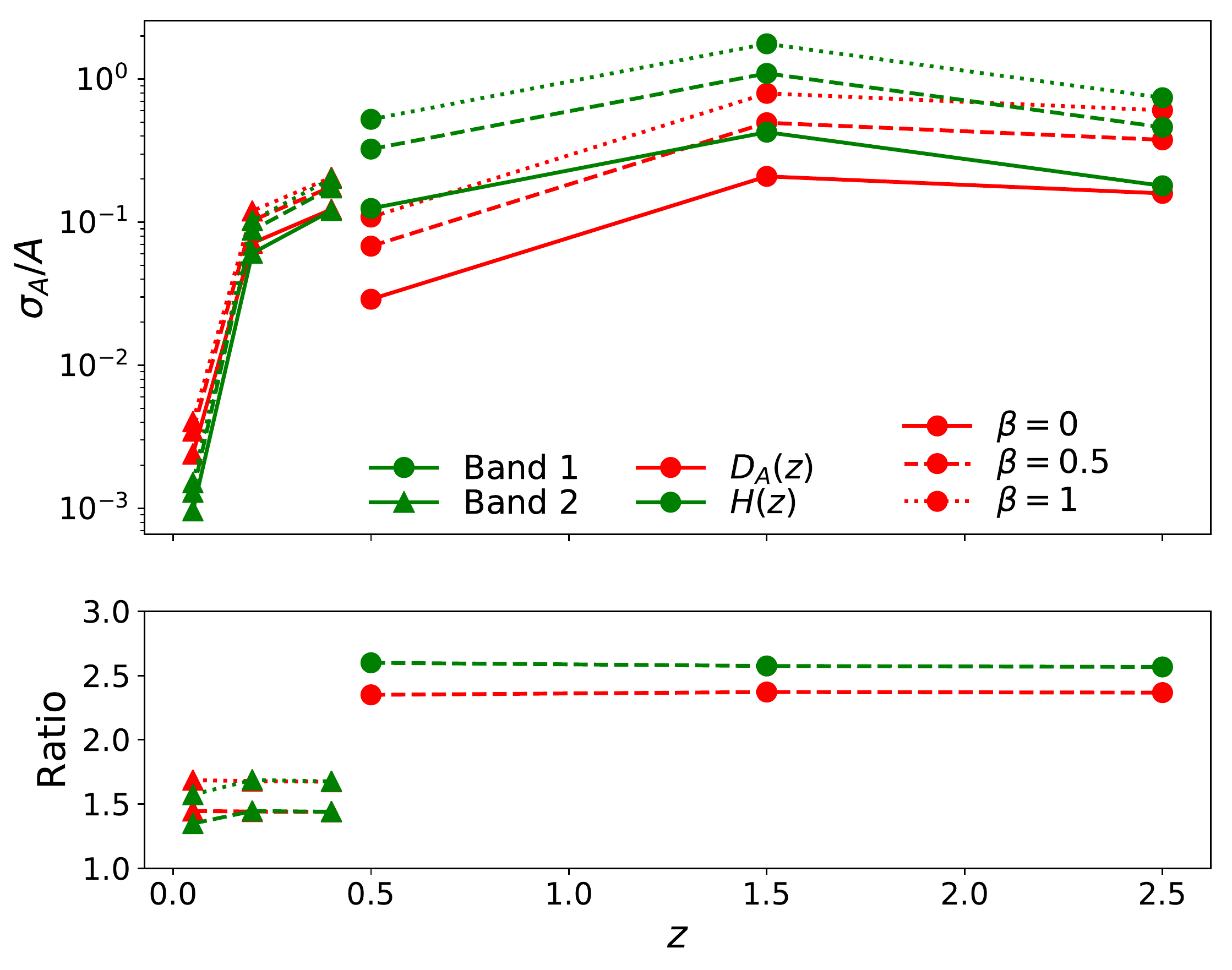}
\includegraphics[width = 0.48\hsize]{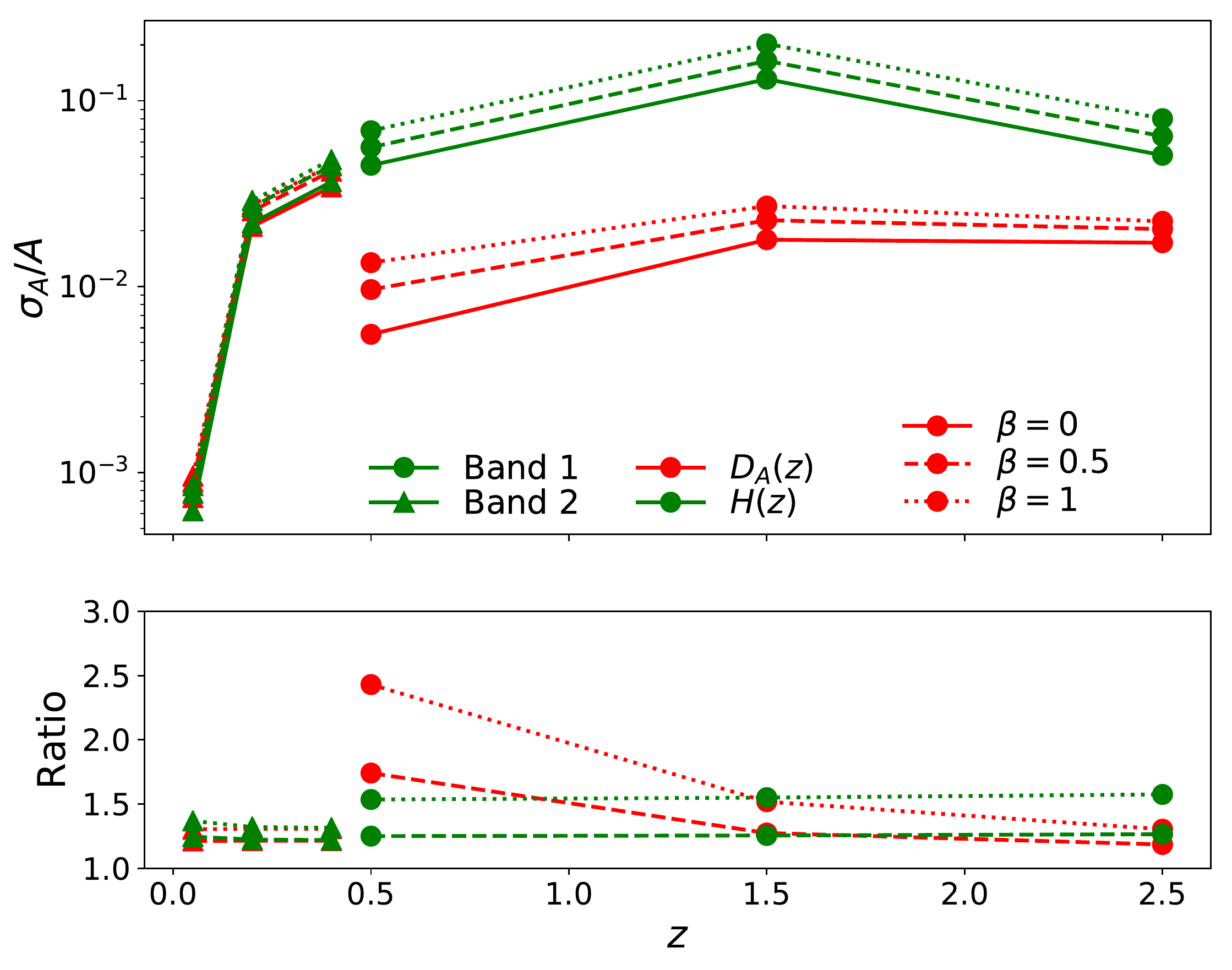}
\end{center}
\caption{The fractional uncertainties of $H(z)$ (\emph{green}) and $D_A(z)$ (\emph{red}) as a function of redshift obtained from SKA Band\,1 (\emph{circle}) and Band\,2 (\emph{triangle}) respectively. The \emph{left} panel is from SKA alone and the \emph{right} panel is from  SKA + \emph{Planck}.  In each case, we consider three scenarios with no $1/f$ noise ($\beta = 0$, \emph{solid}), $\beta = 0.5$ (\emph{dashed}),  and $\beta = 1$ (\emph{dotted}).  The other parameters are set to the baseline values of $v_{\rm t} = 1\,\mathrm{deg\,s^{-1}}$, $f_{\rm knee} = 1\,\mathrm{Hz}$, and $\alpha = 1$.  The \emph{lower} subplot in each panel gives the degradation ratio of the fractional uncertainty with $\beta = 0.5$ ($\beta = 1$) to that with $\beta = 0$.}
\label{fighz}
\end{figure*}
\subsection{Dark energy equation of state}\label{sec:w0-wa}
We now focus on investigating the impact of $1/f$ noise on the dark sector  equation of state parameters $w_0$ and $w_a$. We present the projected joint $w_0-w_a$ $1\,\sigma$  confidence ellipses in Fig.\,\ref{figw0wa}, marginalised over other cosmological parameters.  We consider three cases with effectively no (completely removed) $1/f$ noise  ($\beta = 0$), partially correlated $1/f$ noise ($\beta = 0.5$), and totally uncorrelated $1/f$ noise ($\beta = 1$), with other parameters fixed at the baseline values of $v_{\rm t} = 1\,\mathrm{deg\,s^{-1}}$, $f_k = 1\,\mathrm{Hz}$, and $\alpha = 1$. The \emph{left} panels are for SKA alone and the \emph{right} panels are for SKA+\emph{Planck}, with Band\,1 in \emph{solid} and Band\,2 in \emph{dashed} respectively. 

For $\beta = 0$, the $w_0-w_a$ contour from Band\,1+\emph{Planck} has a similar degenerate direction to that for  Band\,2+\emph{Planck}, although the constraint from Band\,1 alone is tighter than that from Band\,2 alone. This can also be confirmed from Table\,\ref{tabglob} and Fig.\,\ref{fig:hw}, where the addition of IM data to \emph{Planck} breaks the angular diameter distance degeneracy in almost the same way for Band\,1 and Band\,2. Our projected joint $w_0-w_a$ constraints are $\sim50\%$ better than that from \cite{bull16} and the reasons  can be attributed to: i) \cite{bull16} adopted different cosmological parameter set with 11 cosmological parameters while we consider 8 parameters; ii) We adopt the latest SKA configuration with dual polarisation beams while they used  single polarisation; iii) We include  more frequency channels which brings finer information along the \emph{line-of-sight}.

From areas of the contours in  Fig.\,\ref{figw0wa}, we see that  the joint $w_0-w_a$ constraint from Band\,1 alone (\emph{left; solid}) is degraded  by a factor of $\approx2.5$ with $\beta = 0.5$, and $\approx4$ with $\beta = 1$. The contours of Band1+\emph{Planck} (\emph{right; solid}) are less affected by $1/f$ noise, and are degraded by a factor of $\approx1.5$ with $\beta = 0.5$ and $\sim2$ with $\beta = 1$. Band\,2 is also less affected by $1/f$ noise than Band\,1, thanks to its lower redshift where HI signal is stronger.  For Band\,2 alone (\emph{left; dashed}), the joint $w_0-w_a$ constraint is degraded by $\approx1.5$ with $\beta = 0.5$, and $\lesssim2$ with $\beta = 1$. Band\,2+\emph{Planck} (\emph{right; dashed} is much less affected, where the uncertainty  is degraded by less than a factor of $\approx1.3$ even with $\beta = 1$.

The degradation on the joint $w_0-w_a$ plane is consistent with the degradation factor on the power spectrum.  In Sect.\,\ref{sec:ps}, for Band\,1 with $\beta = 0.5$, we see a power spectrum degradation factor of $\approx2$ at $z\lesssim1.5$ and $\ell\lesssim100$, where most of the signal detection comes from, comparable with the factor of $\sim2.5$ degradation on the joint $w_0-w_a$ plane. Similarly, for Band\,2 with $\beta = 0.5$,  the factor of $\approx1.5$ degradation on $w_0-w_a$ is consistent with the power spectrum factor of $\approx1.3$  at $z<0.5$ and $\ell\lesssim200$ where most of the constraining power comes from.

In summary, Band\,1 is more affected by $1/f$ noise than Band\,2 due to observing at higher redshift. With a semi-correlated $1/f$ noise at $\beta = 0.5$,  a degradation by a factor of $\approx1.5$ and $<1.3$ is expected on the joint $w_0-w_a$ plane for Band\,1+\emph{Planck} and Band\,2+\emph{Planck} respectively.

\subsection{Hubble parameter and angular diameter distance}\label{sec:hz}
To study the impact of $1/f$ noise on the expansion history of the Universe, in this section we present projected constraints on the Hubble parameter $H(z)$ and angular diameter distance $D_A(z)$ through parameter transformation described in Sect.\,\ref{sec:fm} at three redshift bins of Band\,1, and  Band\,2 respectively. We calculate the uncertainties, $\sigma_{H(z)}$ (\emph{green}) and $\sigma_{D_A(z)}$ (\emph{red}), relative to the fiducial values of  $H(z)$ and $D_A(z)$ at each  redshift bin, as shown in Fig.\,\ref{fighz} for SKA alone (\emph{left}) and SKA+\emph{Planck} (\emph{right}). In the upper subplot of each panel, we plot the uncertainties for both Band\,1 (\emph{circle}) and Band\,2 (\emph{triangle}) for  three scenarios with  completely removed $1/f$ noise ($\beta = 0$, \emph{solid}), partially correlated $1/f$ noise ($\beta = 0.5$, \emph{dashed}), and completely uncorrelated $1/f$ noise ($\beta = 1$, \emph{dotted}).  The lower subplot gives the corresponding degradation ratio, defined as the ratio of the uncertainty with $\beta = 0.5$ ($\beta = 1$) to that with $\beta = 0$.

It can be seen from the upper subplots of Fig.\,\ref{fighz} that the uncertainties increase with redshift for Band\,2, but reaches its largest value at the middle bin ($z = 1.5$) for Band\,1. The broad peak centred at $z\sim1.5$  is due to the derivatives of $H(z)$ and $D_A(z)$ with respect to $w_0$ and $w_a$ having their peak value at $z \sim 1.5$ and being decreasing afterwards.  Note that our projections on $H(z)$ and $D_A(z)$ are subject to the model assumed in the parameter transformation, which is a very different approach to that in \cite{skaredbook18} who treated $H(z)$ and $D_A(z)$ independently in each bin. This is a very different prior on the allowed values and therefore we expect very different results from the two completely different approaches, even assuming the same instrumental and observing parameters without $1/f$ noise.  We stress that the main point of our analysis is to compare the degradation on cosmological parameter constraints caused by residual $1/f$ noise compared to the case with $\beta = 0$, as opposed to the absolute constraint on the parameters. The degradation ratio between different levels of residual $1/f$ noise, as shown in the lower subplot of Fig.\,\ref{fighz},  is expected to be still representative, were the same methodology adopted as in \cite{skaredbook18}.

Typically, the $1/f$ noise for $\beta = 0.5$ degrades the constraint on $H(z)$ and $D_A(z)$ by a factor of $\approx2.5$ for Band\,1 alone and $\approx1.4$ for Band\,2 alone. A completely uncorrelated $1/f$ noise at $\beta = 1$ will further degrade the results by a factor of  $\approx4$ and  $\approx1.7$ for Band\,1 and Band\,2 respectively. The addition of the \emph{Planck} prior will mitigate the impact of $1/f$ noise so that with $\beta = 0.5$ as an example, the degradation ratio will be reduced to a factor of $\approx1.2$ for both bands.   


\subsection{Growth rate}\label{sec:fz}
\begin{figure}
\begin{center}
\includegraphics[width = 0.9\hsize]{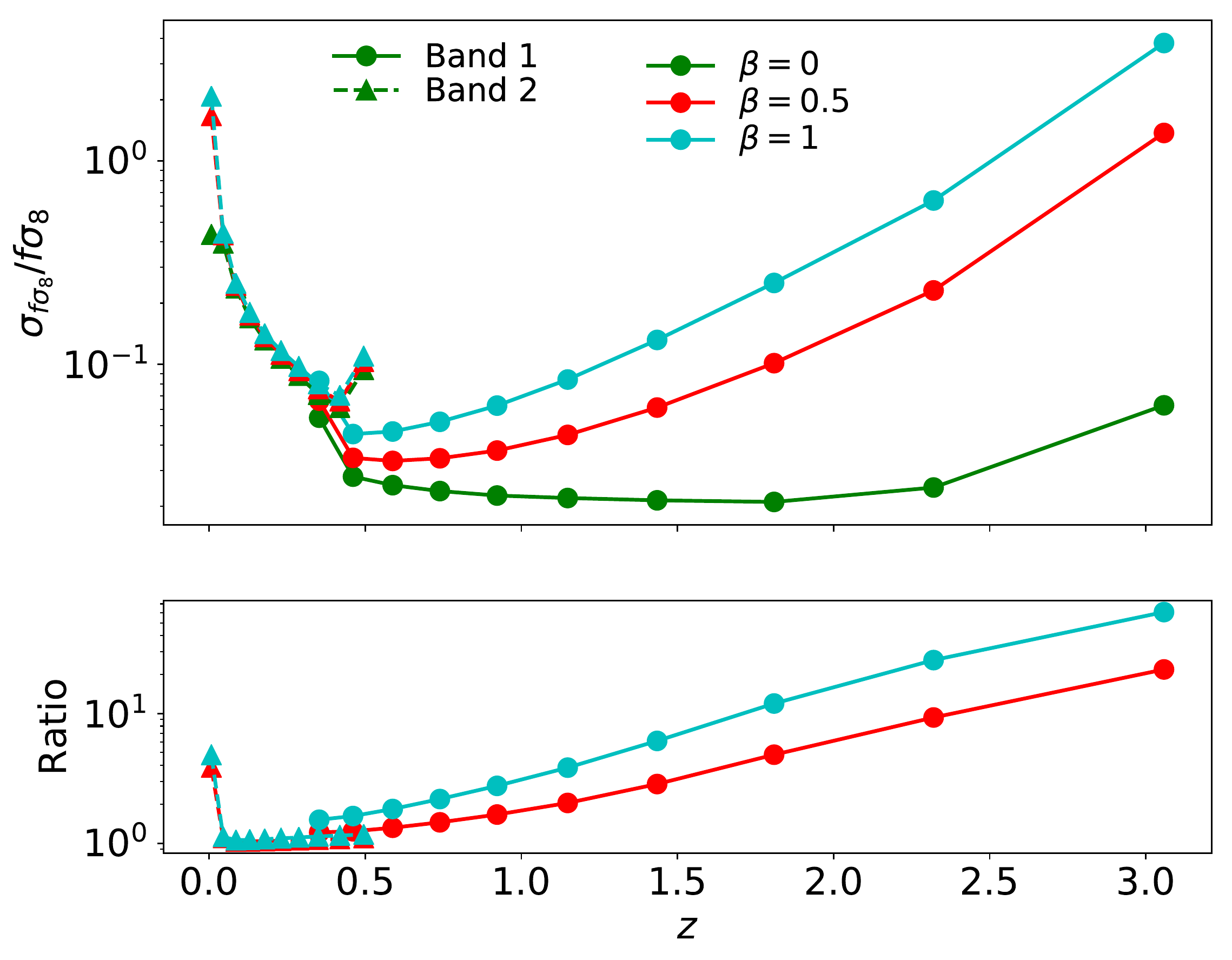}
\end{center}
\caption{The fractional uncertainty of $f\sigma_8(z)$ against redshift obtained from SKA Band\,1 (\emph{circle}) and SKA Band\,2 (\emph{triangle}) respectively. The $1/f$ noise correlation index is set to $\beta = 0$ (effectively no $1/f$ noise; \emph{green}), $\beta = 0.5$ (\emph{red}),  and $\beta = 1$ (\emph{cyan}) respectively.  The other parameters are set to $[v_{\rm t} = 1\,\mathrm{deg\,s^{-1}},\;f_{\rm knee} = 1\,\mathrm{Hz},\;\alpha = 1$]. The \emph{lower} subplot in each panel gives the ratio of the $\beta = 0.5$ and $\beta = 1$ curves over the $\beta= 0$ curve from the \emph{upper} subplot.}
\label{figfz}
\end{figure}

We project constraints on the linear growth rate $f\sigma_8(z) = f(z)\sigma_8D(z)$ in this section as introduced in Sect.\,\ref{sec:fm}. Fig.\,\ref{figfz} shows the fractional uncertainties of $f\sigma_8(z)$ for both Band\,1 and Band\,2, at three $1/f$ noise scenarios of $\beta = 0$, $\beta = 0.5$ and $\beta = 1$. For each band, there are 10 equally spaced frequency  bins of $f\sigma_8(z)$ over the whole band width, where each bin is treated as a free parameter and marginalised over, with all other base parameters fixed. We do not include \emph{Planck} priors in this case. The lower subplot of Fig.\,\ref{figfz} gives the degradation ratio of the fractional uncertainties with $\beta = 0.5$ ($\beta = 1$) to that with $\beta = 0$.

From the upper subplot of Fig.\,\ref{figfz},  our constraints on $f\sigma_8(z) $ are worse than those in \cite{skaredbook18}. This is because while we have a limited $N_{\rm bin} = 35$ and $N_{\rm bin} = 23$ for Band\,1 and Band\,2 respectively, \cite{skaredbook18} calculated the HI power spectrum in the wavenumber,  $k$, space, and thus had a finer resolution and more information along the  \emph{line-of-sight} direction, yielding smaller uncertainties. Again, we clarify that the main point of our paper is to compare results at different residual $1/f$ noise levels. The absolute value of uncertainties subjecting to specific analysis is thus not critical in our case. 

In the lower subplot, with the $1/f$ noise at $\beta = 0.5$,  the uncertainties on $f\sigma_8(z)$ are degraded by  a factor of  $\approx4.8$ and  $\approx1.3$ for Band\,1 and Band\,2 respectively. A larger $\beta$ value will lead to more significant degradation. These are consistent with the results for $H(z)$ and $D_A(z)$.

\section{Conclusions and discussions}\label{dissec}
In this paper, we forecast the constraints on cosmological parameters using the Fisher matrix method for SKA1-MID Band\,1 and Band\,2 in the single dish mode, taking into account the impact of $1/f$ noise using an empirical model from \cite{hdb+18}.

 We begin with the scenario where only thermal noise is  present. With a full HI power spectrum calculation, including cross-correlation frequency bins and redshift-space-distortion contributions, the projected uncertainties are $4\%$ on $w_0$,  $1\%$ on $h$, $2\%$ on $b_{\rm HI}$ using Band\,1+\emph{Planck}, and $3\%$ on $w_0$,  $0.5\%$ on $h$, $2\%$ on $b_{\rm HI}$ using Band\,2+\emph{Planck}. These results would be degraded by $\sim20\%$ on average if one were to exclude contributions from cross-correlation frequency bins. We have tested that excluding the RSD component from the power spectrum calculation will prevent simultaneous measurements on $A_s$ and $b_{\rm HI}$ without the \emph{Planck} prior. We also found that the  parameter constraints improve with increased number of frequency bins thanks to a finer redshift resolution. However, after a certain point, the expensive computing cost from increasing $N_{\rm bin}$ brings little improvement  on the results due to  increased thermal noise with narrower channel width. 

 We study the impact of $1/f$ noise adopting the semi-empirical $1/f$ noise model from \cite{hdb+18}, which quantifies the expected $1/f$ noise level after applying component separation method at the map level. The residual $1/f$ noise spectrum is a function of the frequency correlation index $\beta$, the spectral index $\alpha$, the knee frequency $f_{\rm knee}$, and the telescope slew speed $v_{\rm t}$.  The $1/f$ noise affects the power spectrum detection more significantly at higher redshift due to a weaker HI signal, and smaller scales due to the effect of the beam.  We find that most constraining power comes from  $z\lesssim1$ and $\ell\lesssim100$ at the presence of $1/f$ noise.  A baseline semi-correlated $1/f$ noise at $\beta = 0.5$  degrades the total measured power spectrum  by a factor of $\approx2$ for Band\,1 and a factor of $\approx1.3$ for Band\,2. 

We focus on quantifying the impact of $1/f$ noise on cosmological parameters which IM is very sensitive to, and in particular, the joint constraint on $w_0-w_a$, the Hubble rate $H(z)$, and the angular diameter distance $D_A(z)$.  Typically, the baseline $1/f$ noise at $\beta = 0.5$ degrades these parameter uncertainties by a factor of $\approx2.5$ for Band\,1 and $\approx1.5$ for Band\,2.  This is consistent with the degradation seen in the total power spectrum at the regions where most of the detections come from.  The addition of the \emph{Planck} prior compensates the loss caused by $1/f$ noise and  reduces  the degradation   to $\approx50\%$ for Band\,1+\emph{Planck} and $\approx20\%$ for Band\,2+\emph{Planck}.  The growth rate $f(z)$ is more affected by $1/f$ noise at Band\,1 with a degradation factor of $\approx3$, compared to Band\,2 where there is negligible degradation.  In order to minimise the impact, a minimised $1/f$ noise spectral slope ($\alpha<<2$) and a low knee frequency ($f_{\rm knee}<<0.5\,\mathrm{Hz}$) is critical. A correlation in frequency ($\beta\rightarrow0$) and a large telescope slew speed ($v_{\rm t} \rightarrow 2\,\mathrm{deg\,s^{-1}}$) is also favored.

Our analysis has shown that it is important to control $1/f$ noise for IM experiments.  In particular,  we find that $1/f$ noise can increase the absolute noise level by $\sim2$ orders of magnitude at certain angular scales and redshifts (see Fig.\,\ref{fig:cov1f}). It can  degrade constraints on the growth rate $f(z)$, which is especially valuable in terms of  testing General Relativity \citep[e.g.,][]{jz08, bfs14}. However, even with the representative baseline $1/f$ noise analysed in the paper, IM experiments can still yield decent parameter constraints without too much degradation, although instrumental designs without a proper consideration of $1/f$ noise can potentially result in a much larger degradation factor than quoted here.  We hereby assert that one can no longer ignore the impact of  $1/f$ noise on IM experiments.

We remind the reader that our conclusions are subject to the $1/f$ noise model introduced in \cite{hdb+18}, under  their assumed scan strategy, component separation analysis and perfect calibration without additional systematic errors.  In practice, one may apply additional calibration or filtering techniques  in the time domain to further reduce $1/f$ noise.

\section*{Acknowledgements}

TC acknowledges support from the STFC Innovation Placement scheme, and the Overseas Research Scholarship from the School of Physics and Astronomy, The University of Manchester. RAB ad CD acknowledge support from  an STFC Consolidated Grant (ST/P000649/1). CD also acknowledges an ERC Starting (Consolidator) Grant (no. 307209).


\bibliographystyle{mn2e}
\bibliography{journals,lit} 

\begin{thebibliography}{}

\bibitem[\protect\citeauthoryear{{Alonso}, {Bull}, {Ferreira} et~al.,}{{Alonso}
  et~al.}{2015}]{abf+15}
{Alonso} D.,  {Bull} P.,  {Ferreira} P.~G.,    et~al., 2015, MNRAS, 447, 400

\bibitem[\protect\citeauthoryear{{Anderson}, {Aubourg}, {Bailey}
  et~al.,}{{Anderson} et~al.}{2014}]{aab+14}
{Anderson} L.,  {Aubourg} {\'E}.,  {Bailey} S.,    et~al., 2014, MNRAS, 441, 24

\bibitem[\protect\citeauthoryear{{Ashdown}, {Baccigalupi}, {Balbi}
  et~al.,}{{Ashdown} et~al.}{2007}]{abb+07}
{Ashdown} M.~A.~J.,  {Baccigalupi} C.,  {Balbi} A.,    et~al., 2007, A\&A, 467,
  761

\bibitem[\protect\citeauthoryear{{Asorey}, {Crocce}, {Gazta{\~n}aga}
  et~al.,}{{Asorey} et~al.}{2012}]{acg+12}
{Asorey} J.,  {Crocce} M.,  {Gazta{\~n}aga} E.,    et~al., 2012, MNRAS, 427,
  1891

\bibitem[\protect\citeauthoryear{{Baker}, {Ferreira} \& {Skordis}}{{Baker}
  et~al.}{2014}]{bfs14}
{Baker} T.,  {Ferreira} P.,    {Skordis} C.,  2014, Phys. Rev. D, 89, 024026

\bibitem[\protect\citeauthoryear{{Bandura}, {Addison}, {Amiri}
  et~al.,}{{Bandura} et~al.}{2014}]{baa+14}
{Bandura} K.,  {Addison} G.~E.,  {Amiri} M.,    et~al., 2014, in Society of
  Photo-Optical Instrumentation Engineers (SPIE) Conference Series Vol.~9145 of
  Society of Photo-Optical Instrumentation Engineers (SPIE) Conference Series,
  {Canadian Hydrogen Intensity Mapping Experiment (CHIME) pathfinder}.
p. 914522

\bibitem[\protect\citeauthoryear{{Battye}, {Browne}, {Dickinson}
  et~al.,}{{Battye} et~al.}{2013}]{bbd+13}
{Battye} R.~A.,  {Browne} I. W.~A.,  {Dickinson} C.,    et~al., 2013, MNRAS,
  434, 1239

\bibitem[\protect\citeauthoryear{{Battye}, {Davies} \& {Weller}}{{Battye}
  et~al.}{2004}]{bdw04}
{Battye} R.~A.,  {Davies} R.~D.,    {Weller} J.,  2004, MNRAS, 355, 1339

\bibitem[\protect\citeauthoryear{{Bernal}, {Breysse}, {Gil-Mar{\'\i}n}
  et~al.,}{{Bernal} et~al.}{2019}]{bbg+19}
{Bernal} J.~L.,  {Breysse} P.~C.,  {Gil-Mar{\'\i}n} H.,    et~al., 2019, arXiv
  e-prints:1907.10067

\bibitem[\protect\citeauthoryear{{Bigot-Sazy}, {Dickinson}, {Battye}
  et~al.,}{{Bigot-Sazy} et~al.}{2015}]{bdb+15}
{Bigot-Sazy} M.-A.,  {Dickinson} C.,  {Battye} R.~A.,    et~al., 2015, MNRAS,
  454, 3240

\bibitem[\protect\citeauthoryear{{Bigot-Sazy}, {Ma}, {Battye}
  et~al.,}{{Bigot-Sazy} et~al.}{2016}]{bmb+16}
{Bigot-Sazy} M.-A.,  {Ma} Y.-Z.,  {Battye} R.~A.,    et~al., 2016, in {Qain}
  L.,  {Li} D.,  eds, Frontiers in Radio Astronomy and FAST Early Sciences
  Symposium 2015 Vol.~502 of Astronomical Society of the Pacific Conference
  Series, Hi intensity mapping with fast.
p.~41

\bibitem[\protect\citeauthoryear{{Bonvin} \& {Durrer}}{{Bonvin} \&
  {Durrer}}{2011}]{bd11}
{Bonvin} C.,  {Durrer} R.,  2011, Phys. Rev. D, 84, 063505

\bibitem[\protect\citeauthoryear{{Bowman}, {Cairns}, {Kaplan} et~al.,}{{Bowman}
  et~al.}{2013}]{bck+13}
{Bowman} J.~D.,  {Cairns} I.,  {Kaplan} D.~L.,    et~al., 2013, PASA, 30, e031

\bibitem[\protect\citeauthoryear{{Bull}}{{Bull}}{2016}]{bull16}
{Bull} P.,  2016, ApJ, 817, 26

\bibitem[\protect\citeauthoryear{{Bull}, {Ferreira}, {Patel} et~al.,}{{Bull}
  et~al.}{2015}]{bfp+15}
{Bull} P.,  {Ferreira} P.~G.,  {Patel} P.,    et~al., 2015, ApJ, 803, 21

\bibitem[\protect\citeauthoryear{{Challinor} \& {Lewis}}{{Challinor} \&
  {Lewis}}{2011}]{aa11}
{Challinor} A.,  {Lewis} A.,  2011, Phys. Rev. D, 84, 043516

\bibitem[\protect\citeauthoryear{{Chang}, {Pen}, {Bandura} et~al.,}{{Chang}
  et~al.}{2010}]{cpb+10}
{Chang} T.-C.,  {Pen} U.-L.,  {Bandura} K.,    et~al., 2010, Nature, 466, 463

\bibitem[\protect\citeauthoryear{{Chen}}{{Chen}}{2012}]{chen12}
{Chen} X.,  2012, in International Journal of Modern Physics Conference Series
  Vol.~12 of International Journal of Modern Physics Conference Series, {The
  Tianlai Project: a 21CM Cosmology Experiment}.
pp 256--263

\bibitem[\protect\citeauthoryear{{Chevallier} \& {Polarski}}{{Chevallier} \&
  {Polarski}}{2001}]{cp01}
{Chevallier} M.,  {Polarski} D.,  2001, Int. J. Mod. Phys. D, 10, 213

\bibitem[\protect\citeauthoryear{{Coe}}{{Coe}}{2009}]{coe09}
{Coe} D.,  2009, ArXiv e-prints, ArXiv:0906.4123, instruction for Fisher
  software

\bibitem[\protect\citeauthoryear{{DeBoer}, {Parsons}, {Aguirre}
  et~al.,}{{DeBoer} et~al.}{2017}]{dpa+17}
{DeBoer} D.~R.,  {Parsons} A.~R.,  {Aguirre} J.~E.,    et~al., 2017, PASP, 129,
  045001

\bibitem[\protect\citeauthoryear{{Dodelson}}{{Dodelson}}{2003}]{dod03}
{Dodelson} S.,  2003, {Modern cosmology}.
Academic Press.~ISBN 0-12-219141-2

\bibitem[\protect\citeauthoryear{{Eastwood}, {Anderson}, {Monroe}
  et~al.,}{{Eastwood} et~al.}{2018}]{eam+18}
{Eastwood} M.~W.,  {Anderson} M.~M.,  {Monroe} R.~M.,    et~al., 2018, AJ, 156,
  32

\bibitem[\protect\citeauthoryear{{Fisher}}{{Fisher}}{1920}]{fisher20}
{Fisher} R.~A.,  1920, \mnras, 80, 758

\bibitem[\protect\citeauthoryear{{Golub} \& {van Loan}}{{Golub} \& {van
  Loan}}{1996}]{gv96}
{Golub} G.~H.,  {van Loan} C.~F.,  1996, {Matrix computations}

\bibitem[\protect\citeauthoryear{{Hall} \& {Challinor}}{{Hall} \&
  {Challinor}}{2012}]{hc12}
{Hall} A.~C.,  {Challinor} A.,  2012, MNRAS, 425, 1170

\bibitem[\protect\citeauthoryear{{Hamilton}}{{Hamilton}}{1998}]{hamilton98}
{Hamilton} A.~J.~S.,  1998, in {Hamilton} D.,  ed., The Evolving Universe
  Vol.~231 of Astrophysics and Space Science Library, {Linear Redshift
  Distortions: a Review}.
p.~185

\bibitem[\protect\citeauthoryear{{Harper} \& {Dickinson}}{{Harper} \&
  {Dickinson}}{2018}]{hd18}
{Harper} S.~E.,  {Dickinson} C.,  2018, MNRAS, 479, 2024

\bibitem[\protect\citeauthoryear{{Harper}, {Dickinson}, {Battye}
  et~al.,}{{Harper} et~al.}{2018}]{hdb+18}
{Harper} S.~E.,  {Dickinson} C.,  {Battye} R.~A.,    et~al., 2018, MNRAS, 478,
  2416

\bibitem[\protect\citeauthoryear{{Howlett}, {Lewis}, {Hall} et~al.,}{{Howlett}
  et~al.}{2012}]{hlh+12}
{Howlett} C.,  {Lewis} A.,  {Hall} A.,    et~al., 2012, J. Cosmology
  Astropart.Phys., 4, 027

\bibitem[\protect\citeauthoryear{{Jain} \& {Zhang}}{{Jain} \&
  {Zhang}}{2008}]{jz08}
{Jain} B.,  {Zhang} P.,  2008, Phys. Rev. D, 78, 063503

\bibitem[\protect\citeauthoryear{{Kovetz}, {Viero}, {Lidz} et~al.,}{{Kovetz}
  et~al.}{2017}]{kvl+17}
{Kovetz} E.~D.,  {Viero} M.~P.,  {Lidz} A.,    et~al., 2017, arXiv e-prints:
  1709.09066

\bibitem[\protect\citeauthoryear{{Lewis} \& {Bridle}}{{Lewis} \&
  {Bridle}}{2002}]{lb02}
{Lewis} A.,  {Bridle} S.,  2002, Phys. Rev. D, 66, 103511

\bibitem[\protect\citeauthoryear{{Linder}}{{Linder}}{2003}]{lin03}
{Linder} E.~V.,  2003, Phys. Rev. Lett., 90, 091301

\bibitem[\protect\citeauthoryear{{Madau}, {Meiksin} \& {Rees}}{{Madau}
  et~al.}{1997}]{mmr97}
{Madau} P.,  {Meiksin} A.,    {Rees} M.~J.,  1997, ApJ, 475, 429

\bibitem[\protect\citeauthoryear{{Masui}, {Switzer}, {Banavar} et~al.,}{{Masui}
  et~al.}{2013}]{msb+13}
{Masui} K.~W.,  {Switzer} E.~R.,  {Banavar} N.,    et~al., 2013, ApJ, 763, L20

\bibitem[\protect\citeauthoryear{{Nan}, {Li}, {Jin} et~al.,}{{Nan}
  et~al.}{2011}]{nlj+11}
{Nan} R.,  {Li} D.,  {Jin} C.,    et~al., 2011, Int. J. Mod. Phys. D, 20, 989

\bibitem[\protect\citeauthoryear{{Newburgh}, {Bandura}, {Bucher}
  et~al.,}{{Newburgh} et~al.}{2016}]{nbb+16}
{Newburgh} L.~B.,  {Bandura} K.,  {Bucher} M.~A.,    et~al., 2016, in
  Ground-based and Airborne Telescopes VI Vol.~9906 of Proc. SPIE, {HIRAX: a
  probe of dark energy and radio transients}.
p. 99065X

\bibitem[\protect\citeauthoryear{Nyquist}{Nyquist}{1928}]{Nyquist1928}
Nyquist H.,  1928, Physical Review, 32, 110

\bibitem[\protect\citeauthoryear{{Olivari}, {Dickinson}, {Battye}
  et~al.,}{{Olivari} et~al.}{2018}]{odb+18}
{Olivari} L.~C.,  {Dickinson} C.,  {Battye} R.~A.,    et~al., 2018, MNRAS, 473,
  4242

\bibitem[\protect\citeauthoryear{{Olivari}, {Remazeilles} \&
  {Dickinson}}{{Olivari} et~al.}{2016}]{ord16}
{Olivari} L.~C.,  {Remazeilles} M.,    {Dickinson} C.,  2016, MNRAS, 456, 2749

\bibitem[\protect\citeauthoryear{{Padmanabhan}, {Choudhury} \&
  {Refregier}}{{Padmanabhan} et~al.}{2015}]{pcr15}
{Padmanabhan} H.,  {Choudhury} T.~R.,    {Refregier} A.,  2015, MNRAS, 447,
  3745

\bibitem[\protect\citeauthoryear{{Parsons}, {Backer}, {Foster}
  et~al.,}{{Parsons} et~al.}{2010}]{pbf+10}
{Parsons} A.~R.,  {Backer} D.~C.,  {Foster} G.~S.,    et~al., 2010, AJ, 139,
  1468

\bibitem[\protect\citeauthoryear{{Patil}, {Yatawatta}, {Koopmans}
  et~al.,}{{Patil} et~al.}{2017}]{pyk+17}
{Patil} A.~H.,  {Yatawatta} S.,  {Koopmans} L.~V.~E.,    et~al., 2017, ApJ,
  838, 65

\bibitem[\protect\citeauthoryear{{Peterson}, {Bandura} \& {Pen}}{{Peterson}
  et~al.}{2006}]{pbp06}
{Peterson} J.~B.,  {Bandura} K.,    {Pen} U.~L.,  2006, ArXiv Astrophysics
  e-prints, ArXiv:0606104, Presented at Moriond Cosmology 2006

\bibitem[\protect\citeauthoryear{{Planck Collaboration VI}}{{Planck
  Collaboration VI}}{2018}]{planckVI18}
{Planck Collaboration VI} 2018, ArXiv e-prints, ArXiv:1807.06209, submitted to
  A\&A

\bibitem[\protect\citeauthoryear{{Planck Collaboration XI}}{{Planck
  Collaboration XI}}{2016}]{planckXI15}
{Planck Collaboration XI} 2016, A\&A, 594, A11

\bibitem[\protect\citeauthoryear{{Santos}, {Bull}, {Alonso} et~al.,}{{Santos}
  et~al.}{2015}]{sba+15}
{Santos} M.,  {Bull} P.,  {Alonso} D.,    et~al., 2015, Advancing Astrophysics
  with the Square Kilometre Array (AASKA14), p.~19

\bibitem[\protect\citeauthoryear{{Seiffert}, {Mennella}, {Burigana}
  et~al.,}{{Seiffert} et~al.}{2002}]{smb+02}
{Seiffert} M.,  {Mennella} A.,  {Burigana} C.,    et~al., 2002, A\&A, 391, 1185

\bibitem[\protect\citeauthoryear{{SKA Red Book}.}{{SKA Red
  Book}.}{2018}]{skaredbook18}
{SKA Red Book}. 2018, arXiv e-prints, arXiv:1811.02743

\bibitem[\protect\citeauthoryear{{Switzer}, {Masui}, {Bandura}
  et~al.,}{{Switzer} et~al.}{2013}]{smb+13}
{Switzer} E.~R.,  {Masui} K.~W.,  {Bandura} K.,    et~al., 2013, MNRAS, 434,
  L46

\bibitem[\protect\citeauthoryear{{van Haarlem}, {Wise}, {Gunst} et~al.,}{{van
  Haarlem} et~al.}{2013}]{hwg+13}
{van Haarlem} M.~P.,  {Wise} M.~W.,  {Gunst} A.~W.,    et~al., 2013, A\&A, 556,
  A2

\bibitem[\protect\citeauthoryear{{Weinberg}, {Mortonson}, {Eisenstein}
  et~al.,}{{Weinberg} et~al.}{2013}]{wme+13}
{Weinberg} D.~H.,  {Mortonson} M.~J.,  {Eisenstein} D.~J.,    et~al., 2013,
  Phys. Rep., 530, 87

\bibitem[\protect\citeauthoryear{{Wilson}, {Rohlfs} \&
  {Hüttemeister}}{{Wilson} et~al.}{2009}]{wrh09}
{Wilson} T.~L.,  {Rohlfs} K.,    {Hüttemeister} S.,  2009, Tools of Radio
  Astronomy.
Springer Berlin Heidelberg

\bibitem[\protect\citeauthoryear{{Wolz}, {Abdalla}, {Blake} et~al.,}{{Wolz}
  et~al.}{2014}]{wab+14}
{Wolz} L.,  {Abdalla} F.~B.,  {Blake} C.,    et~al., 2014, MNRAS, 441, 3271

\end{thebibliography}


\bsp	
\label{lastpage}
\end{document}